\pdfoutput=1
\documentclass[cernpreprint, atlasdraft=false, texlive=2023, UKenglish, texmf, orcidlogo]{atlasdoc}
 
\usepackage{atlaspackage}
\usepackage{atlasbiblatex}
 
\usepackage{atlasphysics}
 
\usepackage[utf8]{inputenc}
 
\usepackage{booktabs}
\usepackage{multirow}
\usepackage{pifont}
\usepackage{MnSymbol}
\usepackage{numprint}
\usepackage{array, siunitx}
\usepackage{cleveref}
 
\usepackage{xcolor}
\definecolor{hh-dark-pink}{HTML}{F2385A}
\definecolor{hh-dark-blue}{HTML}{343844}
\definecolor{hh-turquoise}{HTML}{36B1BF}
\definecolor{hh-light-turquoise}{HTML}{4AD9D9}
\definecolor{hh-off-white}{HTML}{E9F1DF}
\definecolor{hh-dark-yellow}{HTML}{FDC536}

\graphicspath{{logos/}{figures/}}
 
\usepackage{ANA-HDBS-2019-04-PAPER-defs}
 
\usepackage{cancel}
 
\usepackage{etoolbox}
\providetoggle{conf}
\settoggle{conf}{false}
 
\newenvironment{DIFnomarkup}{}{}

\AtlasTitle{Search for non-resonant Higgs boson pair production in final states with leptons, taus, and photons in $pp$\ collisions at $\sqrt{s} = 13\,$TeV with the ATLAS detector}

\AtlasAbstract{
A search is presented for non-resonant Higgs boson pair production,
targeting the $bbZZ$, 4$V$\ ($V$\ = $W$\ or $Z$), $VV\tau\tau$, 4$\tau$,
$\gamma\gamma VV$\ and $\gamma\gamma\tau\tau$\ decay channels.  Events
are categorised based on the multiplicity of light charged leptons
(electrons or muons), hadronically decaying tau leptons, and photons.
The search is based on a data sample of proton--proton collisions at
$\sqrt{s}$~=~13\,TeV recorded with the ATLAS detector during Run 2 of
the Large Hadron Collider, corresponding to an integrated luminosity
of 140\,\mbox{fb\(^{-1}\)}.  No evidence of the signal is found and
the observed (expected) upper limit on the cross-section for
non-resonant Higgs boson pair production is determined to be \muHHobs\
(\muHHexpzero) times the Standard Model predicted cross-section at
95\% confidence level under the background-only hypothesis.
The observed (expected) constraints on the $HHH$\ coupling
modifier, $\kappa_{\lambda}$, are determined to be $\klobsdown < \kappa_{\lambda}\
< \klobsup$\ ($\klexpdown < \kappa_{\lambda}\ < \klexpup$) at 95\% confidence level,
assuming the Standard Model for the expected limits and that new
physics would only affect $\kappa_{\lambda}$.
}

\AtlasRefCode{HDBS-2019-04}
 
\PreprintIdNumber{CERN-EP-2024-131}
 
\AtlasDate{\today}

\AtlasJournalRef{\JHEP 08 (2024) 164}
\AtlasDOI{10.1007/JHEP08(2024)164}

\hypersetup{pdftitle={ATLAS document},pdfauthor={The ATLAS Collaboration}}
 
\begin{document}
 
\maketitle
 
\tableofcontents

\section{Introduction}

Since the discovery of the Higgs boson by the ATLAS and CMS
Collaborations in 2012~\cite{HIGG-2012-27,CMS-HIG-12-028}, a major
focus in particle physics has been understanding how the Higgs boson
interacts with other particles.  Tremendous progress has been made in
determining the strength of the Higgs boson's couplings to fermions
and vector bosons~\cite{HIGG-2021-23,CMS-HIG-22-001}, but its
self-interaction has yet to be established.  Measurements of the Higgs
self-coupling are an essential component of understanding electroweak
symmetry breaking and are a sensitive probe for new physics.  Many
models for new physics predict the existence of additional particles,
the presence of which would lead to deviations from the Standard Model
(SM) prediction for the strength of the Higgs self-coupling.
Moreover, the nature of the electroweak phase transition, when the
electromagnetic and weak forces differentiated as the universe cooled
down after the Big Bang, is still unknown.  The SM predicts a smooth
continuous cross-over from one phase to the other, but a first-order
phase transition is needed in most models in order to accommodate
phenomena like baryonic asymmetry with baryogenesis.  New particles or
fields that interact with the Higgs boson are required to accommodate
the needed first-order phase transition, and this in turn may lead to
a large modification ($\mathcal{O}(1)$\ times the SM prediction) to
the Higgs self-coupling~\cite{Horn,Reichert,Noble:2007kk}.  Some
inflation models require that the Higgs boson couples to gravity,
which in turn modifies the shape of the Higgs
potential~\cite{Bezrukov:2007ep}.  Measurements of the Higgs
self-coupling can provide important information to constrain such
models.  In addition to providing information about the formation of
the universe, Higgs self-coupling measurements can also proffer
insight into its stability~\cite{Markkanen}\ and eventual fate.
 
\begin{figure}[!htb]
\centering
\subfloat[]{ \includegraphics[width=0.4\textwidth]{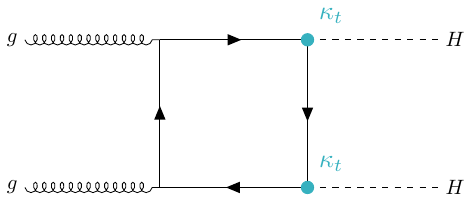}
}
\subfloat[]{ \includegraphics[width=0.4\textwidth]{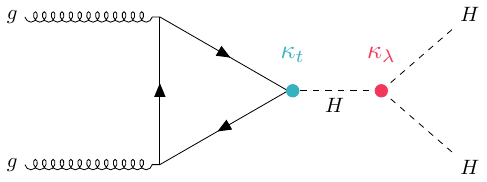}
}
\caption{Leading-order diagrams for gluon--gluon fusion \HH\ production,
via (a) the top-quark box, and (b) the self-interaction `triangle' modes.  The
$ttH$\ and $HHH$\ coupling strength modifiers are denoted as \kt\ and
\kl, respectively.
}
\label{fig:ggF_HH}
\end{figure}

The most natural way to probe the Higgs self-interaction is via
searches for Higgs boson pair production, \HH.  At the LHC the
dominant \HH\ production mode in the SM is gluon--gluon fusion (\ggF).
The leading-order (LO) Feynman diagrams for this process are shown in
Figure~\ref{fig:ggF_HH}.  The \ggF\ cross-section, for a Higgs boson
mass of $\mH\ = 125$\,\GeV, calculated at next-to-next-to-leading-order
(NNLO) accuracy in the finite top-quark mass approximation, is
$\sigma_{HH}(\ggF) = 31.1^{+6.7\%}_{-23.2\%}$~fb~\cite{Dawson:1998py,Borowka:2016ehy,Baglio:2018lrj,deFlorian:2013jea,Shao:2013bz,deFlorian:2015moa,Grazzini:2018bsd,Baglio:2020wgt}.
The two \ggF\ production modes shown in
Figure~\ref{fig:ggF_HH}\ interfere with each other destructively in
the SM\@.  The cross-section of the $pp\rightarrow HH$ process and shape
of the \mHH\ distribution both change as the strength of the Higgs
self-coupling relative to the SM prediction (denoted by \kl\
= \lHHH/$\lambda_{SM}$) is varied.
 
\begin{figure}[!htb]
\centering
\subfloat[]{ \includegraphics[width=0.25\textwidth]{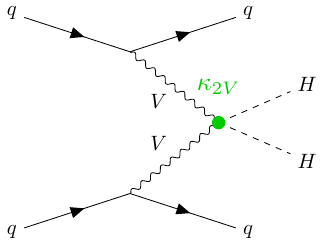}
}
\subfloat[]{ \includegraphics[width=0.25\textwidth]{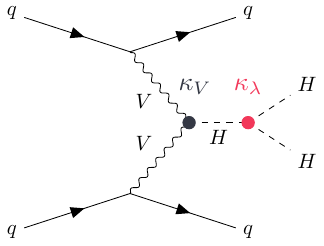}
}
\subfloat[]{ \includegraphics[width=0.25\textwidth]{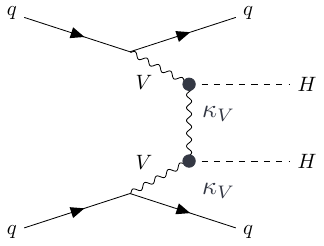}
}
\caption{Leading-order diagrams for VBF \HH\ production containing
(a) the $HHVV$\ vertex, (b) the trilinear $HHH$\ vertex, and (c) via the
$VVHH$\ production mode. The $HHVV$\ and $HVV$\ coupling strength modifiers
are denoted by \kVV\ and $\kappa_{V}$, respectively.
}
\label{fig:VBF_HH}
\end{figure}

The vector-boson fusion (VBF) \HH\ process provides a sub-leading
source of \HH\ production in the SM, and has a cross-section of
$1.73 \pm2.1\%$~fb, calculated at next-to-next-to-next-to-leading
order (N$^3$LO) with $\mH =
125$\,\GeV~\cite{Baglio:2012np,Frederix:2014hta,Ling:2014sne,Dreyer:2018rfu,Dreyer:2018qbw,Dreyer:2020xaj}.
The VBF production mode provides sensitivity to the $VVHH$ coupling
(where the coupling strength with respect to the SM prediction is
denoted as \kVV), as well as additional sensitivity to the Higgs
self-coupling, as shown in Figure~\ref{fig:VBF_HH}.  Both the
gluon--gluon fusion and VBF production modes of Higgs boson pairs are
considered as signal in this paper.  Other production modes have lower
cross-sections and are neglected.
 
Many searches for \HH\ production have been made by both the ATLAS and
CMS Collaborations.  A statistical combination of ATLAS results in
the \HHbbyy~\cite{HDBS-2018-34},
\HHbbtautau~\cite{HDBS-2018-40}, and \HHbbbb~\cite{HDBS-2019-29}
\footnote{Where unspecified, charge conjugation is implied and
the notations $\tau\tau$, $WW$\ etc. are used in place of
$\tau^{+}\tau^{-}$, $W^{+}W^{-}$ etc.}
channels using the full Run 2 data set (up to 140\,\ifb of data collected
during 2015--2018 with centre-of-mass energy \sqrts = 13\,\TeV) sets
an observed (expected) upper limit on the Higgs boson pair production
cross-section at 2.4 (2.9) times the SM prediction at 95\% confidence
level (CL)~\cite{HDBS-2022-03}.  These results are further combined
with precision measurements of single Higgs boson production to
constrain the self-coupling strength modifier to be within the range
of $-0.4 \leq \kl \leq 6.3$\ ($-1.9 \leq \kl \leq 7.6$\ expected) at
95\% CL\@.  More recent results improve the performance of the
individual \HHbbyy~\cite{HDBS-2021-10}\
and \HHbbtautau~\cite{HDBS-2019-27}\ channels, and probe the \HH\
process in final states with two $b$-jets, two light leptons ($\ell =
e/\mu$) and missing transverse momentum
(\met)~\cite{HDBS-2019-02}, but the aforementioned combination
continues to set the overall strongest limits on the \HH\
cross-section and self-coupling strength.  An analysis exploring the
VBF \HHbbbb\ channel in boosted toplogies~\cite{HDBS-2022-02}\
improves upon the \kVV\ coupling constraints of
Ref.~\cite{HDBS-2022-03}.
 
The CMS Collaboration achieves similar sensitivity to the ATLAS
results in a combination of results~\cite{CMS-HIG-22-001} from
analyses of the \HHbbyy~\cite{CMS-HIG-19-018},
\HHbbtautau~\cite{CMS-HIG-20-010},
\HHbbbb~\cite{CMS-HIG-20-005,CMS-B2G-22-003}\ and
$HH\rightarrow bbZZ(ZZ\rightarrow 4\ell)$~\cite{CMS-HIG-20-004}\ decay
channels, and a `multilepton' analysis
covering the $HH\rightarrow 4W$, $WW\tau\tau$, and $4\tau$\ decay
modes in final states with two, three or four light leptons or
hadronic taus~\cite{CMS-HIG-21-002}.  This combination
sets an observed (expected) upper limit on the \HH\ cross-section
of 3.4 (2.5) times the SM prediction and constrains the self-coupling
strength to $-1.24 \leq \kl \leq 6.49$.
The analysis probing the $HH\rightarrow bbZZ(ZZ\rightarrow
4\ell)$\ decay mode sets an observed (expected) limit on the
cross-section of 32 (40) times the SM prediction and constrains the
Higgs boson self-coupling strength to be $-8.8 \leq \kl \leq 13.4$\
($-9.8 \leq \kl \leq 15.0$), all at 95\% CL.  The multilepton analysis
sets an observed (expected) upper limit on the cross-section of 21.3
(19.4) times the SM prediction and constrains the Higgs boson
self-coupling strength to be $-6.9 \leq \kl \leq 11.1$\
($-6.9 \leq \kl \leq 11.7$) at 95\% CL.
 
The analysis described here provides a complementary way to probe
Higgs boson pair production by targeting
production of the \HH\ process in final states with multiple light
leptons and hadronic taus, and in diphoton final states with one or
two additional light leptons and/or hadronic taus (\tauh).  A
visualisation of the final states covered in this analysis is shown in
Figure~\ref{fig:channel_map}.  The analysis is designed to select
events from \HH\ decays where $H \rightarrow WW$, $ZZ$, $\tau\tau$,
and $\gamma\gamma$.  The $HH\rightarrow bbZZ$\ decay mode, with both
the $Z$\ bosons undergoing a decay to light leptons, is also analysed
in a dedicated search.  Channels including an
$H\rightarrow\gamma\gamma$\ decay are classified as the diphoton plus
multilepton channels (`\yyML') while those without photons are
classified as multilepton (`ML') channels.  This is the first time
these \HH\ decay channels are explored in a multilepton analysis in
ATLAS\@.  The event selections are orthogonal by construction with
those used in the ATLAS analyses of
the \bbyy~\cite{HDBS-2018-34,HDBS-2021-10},
\bbtautau~\cite{HDBS-2018-40,HDBS-2019-27},
\bbbb~\cite{HDBS-2019-29,HDBS-2022-02}, and
\bbll~\cite{HDBS-2019-02}\ \HH\ decay channels. Boosted decision
trees (BDTs) are used to enhance signal to background separation.
Upper limits are set on the \HH\ signal strength, \muHH\ (defined as
the ratio of the \HH\ production cross-section, including only the
\ggF\ and \VBF\ processes, to its SM prediction of 32.8\,fb), and the
coupling strength modifiers \kl\ and \kVV, all at 95\% CL.
 
\begin{figure}[ht]
\centering
\includegraphics[width=0.9\textwidth]{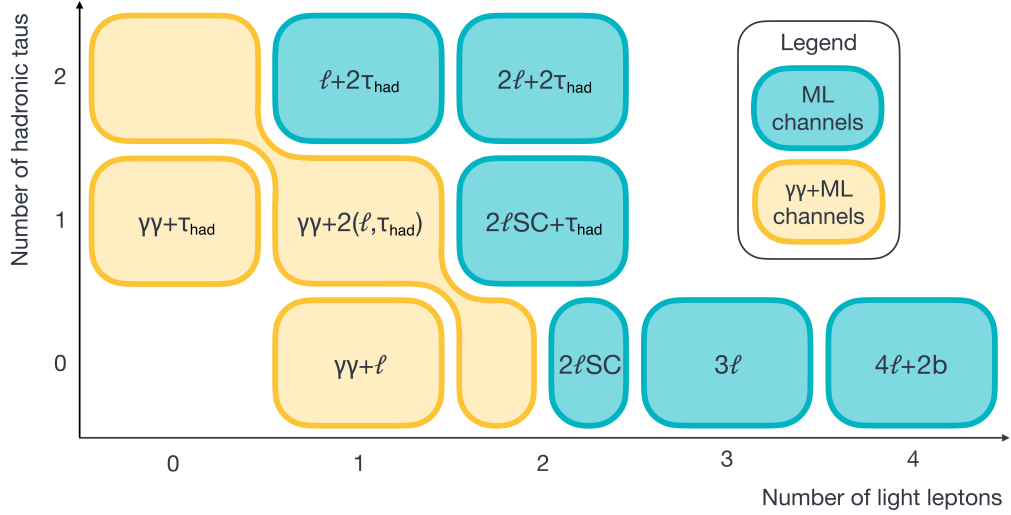}
 
\caption{A visualisation of the different final states included in
this analysis.  The diphoton plus multilepton channels (`\yyML\
channels') are shown in the lighter yellow boxes and channels with light leptons
and hadronic taus (`ML channels') are indicated by the darker turquoise boxes.
`SC' indicates that the two leptons are required to have the
same charge.  The two hadronic taus in the \lltautau\
and \ltautau\ channels are required to have opposite charge
(`OC'), as are the two light leptons in the \lltautau\ channel.
The \yyll\ channel requires the presence of two OC light leptons or
hadronic taus in addition to the two photons, i.e. encompassing
$\gamma\gamma + \ell\ell$, $\gamma\gamma + \ell\tauh$, and
$\gamma\gamma + \tauh\tauh$\ signatures.}
\label{fig:channel_map}
\end{figure}


\section{ATLAS detector}

\newcommand{\AtlasCoordFootnote}{ ATLAS uses a right-handed coordinate
system with its origin at the nominal interaction point (IP) in the
centre of the detector and the \(z\)-axis along the beam pipe. The
\(x\)-axis points from the IP to the centre of the LHC ring, and the
\(y\)-axis points upwards. Polar coordinates \((r,\phi)\) are used
in the transverse plane, \(\phi\) being the azimuthal angle around
the \(z\)-axis. The pseudorapidity is defined in terms of the polar
angle \(\theta\) as \(\eta = -\ln \tan(\theta/2)\) and is equal to
the rapidity $ y = \frac{1}{2} \ln \left( \frac{E + p_z c}{E - p_z
c} \right) $ in the relativistic limit. Angular distance is
measured in units of \(\Delta R \equiv \sqrt{(\Delta y)^{2} +
(\Delta\phi)^{2}}\).}
 
The ATLAS detector~\cite{PERF-2007-01} at the LHC~\cite{Evans:2008zzb}
covers nearly the entire solid angle around the collision
point.\footnote{\AtlasCoordFootnote} It consists of an inner tracking
detector surrounded by a thin superconducting solenoid,
electromagnetic and hadron calorimeters, and a muon spectrometer
incorporating three large superconducting air-core toroidal magnets.
 
The inner-detector system is immersed in a \qty{2}{\tesla} axial
magnetic field and provides charged-particle tracking in the range
\(|\eta| < 2.5\).  The high-granularity silicon pixel detector covers
the vertex region and typically provides four measurements per track,
the first hit normally being in the insertable B-layer (IBL) installed
before Run~2~\cite{ATLAS-TDR-19_copy,PIX-2018-001}.  It is followed by
the silicon microstrip tracker (SCT), which usually provides eight
measurements per track.  These silicon detectors are complemented by
the transition radiation tracker (TRT), which enables radially
extended track reconstruction up to \(|\eta| = 2.0\).  The TRT also
provides electron identification information based on the fraction of
hits (typically 30 in total) above a higher energy-deposit threshold
corresponding to transition radiation.
 
The calorimeter system covers the pseudorapidity range \(|\eta| <
4.9\).  Within the region \(|\eta|< 3.2\), electromagnetic calorimetry
is provided by barrel and endcap high-granularity lead/liquid-argon
(LAr) calorimeters, with an additional thin LAr presampler covering
\(|\eta| < 1.8\) to correct for energy loss in material upstream of
the calorimeters.  Hadron calorimetry is provided by the
steel/scintillator-tile calorimeter, segmented into three barrel
structures within \(|\eta| < 1.7\), and two copper/LAr hadron endcap
calorimeters.  The solid angle coverage is completed with forward
copper/LAr and tungsten/LAr calorimeter modules optimised for
electromagnetic and hadronic energy measurements respectively.
 
The muon spectrometer (MS) comprises separate trigger and
high-precision tracking chambers measuring the deflection of muons in
a magnetic field generated by the superconducting air-core toroidal
magnets.  The field integral of the toroids ranges between \num{2.0}
and \qty{6.0}{\tesla\metre} across most of the detector.  Three layers
of precision chambers, each consisting of layers of monitored drift
tubes, cover the region \(|\eta| < 2.7\), complemented by
cathode-strip chambers in the forward region, where the background is
highest.  The muon trigger system covers the range \(|\eta| < 2.4\)
with resistive-plate chambers in the barrel, and thin-gap chambers in
the endcap regions.
 
Interesting events are selected by the first-level trigger system
implemented in custom hardware, followed by selections made by
algorithms implemented in software in the high-level
trigger~\cite{TRIG-2016-01}.  The first-level trigger accepts events
from the \qty{40}{\MHz} bunch crossings at a rate below
\qty{100}{\kHz}, which the high-level trigger further reduces in order
to record events to disk at about \qty{1}{\kHz}.
 
An extensive software suite~\cite{ATL-SOFT-PUB-2021-001} is used in
data simulation, in the reconstruction and analysis of real and
simulated data, in detector operations, and in the trigger and data
acquisition systems of the experiment.


\section{Data and simulated event samples}

\subsection{Data sample}
 
The analysis is performed using proton--proton ($pp$) collision data with
\sqrts~=~13\,\TeV\ collected during the LHC Run~2.  The number of
$pp$\ interactions per bunch crossing (pile-up) in this
data set ranges from about 8 to 70, with an average of 34. After
applying data quality requirements~\cite{DAPR-2018-01} the full
data set has an integrated luminosity of $140.1 \pm 1.2$\,\ifb~\cite{DAPR-2021-01,LUCID2}.
The trigger requirements are discussed in Section~\ref{sec:events}.
 
\subsection{Simulated event samples}
\label{subsec:mc}

Monte Carlo (MC) simulation is used for the modelling of signal
events and most background processes. All generated events are
processed through a simulation~\cite{SOFT-2010-01} of the ATLAS
detector geometry and response using
\GEANT~\cite{Agostinelli:2002hh}, and through the same
reconstruction software as the data.  Corrections are applied to the
simulated events so that the particle candidates' selection
efficiencies, energy scales and resolutions match those
determined from data control samples.  The samples of simulated events
are normalised to the corresponding predicted cross-sections, computed
to the highest order available in perturbation theory.  The mass of
the top quark and Higgs boson are set to $m_{t}$\ = 172.5\,\GeV
and $m_{H}$\ = 125\,\GeV, respectively.
 
The effect of \pileup\ is modelled by overlaying the simulated
hard-scattering event with inelastic $pp$\ events
generated with \PYTHIA[8.186]~\cite{Sjostrand:2007gs} using the
\NNPDF[2.3lo] set of parton distribution functions
(PDF)~\cite{Ball:2012cx_copy} and the A3 set of tuned
parameters~\cite{ATL-PHYS-PUB-2016-017}.  The simulated events
are weighted to reproduce the distribution of the average number of
interactions per bunch crossing (\(\left<\mu \right>\)) observed in
the data. The \(\left<\mu \right>\) value in data is rescaled by a
factor of \(1.03\pm 0.04\) to improve agreement between data and
simulation in the visible inelastic $pp$\ cross-section~\cite{STDM-2015-05}.
 
The \ggF~\HH\ signal process is simulated using the \POWHEGBOX[v2]
generator~\cite{Alioli:2010xd,Nason:2009ai} at next-to-leading-order
(NLO), including finite top-quark-mass effects~\cite{Borowka:2016ehy},
using the \PDFforLHC[15]~\cite{Butterworth:2015oua}\ PDF set.  Parton
showers and hadronisation are simulated with
\PYTHIA[8.244]~\cite{Sjostrand:2014zea} with the A14 set of tuned
parton shower parameters~\cite{ATL-PHYS-PUB-2014-021} and the
\NNPDF[2.3lo] PDF set.  Signal samples for the ggF
process are generated explicitly for coupling modifier values of
$\kl=1$ and 10.  A reweighting method is used to obtain a \ggF\
signal sample at other \kl\ values by performing a linear combination
of independent generator-level samples at three different \kl\ values
($\kl=0,1,$ and $20$), following the method described in
Ref.~\cite{ATL-PHYS-PUB-2019-007}. Scale factors are derived as a
function of \kl\ in bins of the generator-level invariant mass of the
\HH\ system and applied to the simulated \ggF, $\kl=1$\ sample.
The \ggF, $\kl=10$\ signal sample is used to validate the derived scale
factors.  This sample and the signal sample obtained from
the reweighting method are found to agree within their statistical
precision.  For the reweighted \ggF\ signal,
the NNLO cross-section as a function of \kl\ is taken from
Ref.~\cite{Grazzini:2018bsd}.  To assess parton showering
uncertainties, alternative ggF samples are simulated using the
\POWHEGBOX[v2] generator at NLO with the \PDFforLHC[15] PDF set,
interfaced to \HERWIG[7.1.6]~\cite{Bahr:2008pv} for parton showering
and hadronisation using the \HERWIG[7.1] default set of tuned
parameters~\cite{Gieseke:2012ft} and \MMHT[lo] PDF
set~\cite{Harland-Lang:2014zoa} for parton showering and hadronisation. 
 
The VBF \HH\ signal process is simulated using
\MGNLO[2.7.3]~\cite{Alwall:2014hca} at LO with the \NNPDF[3.0nlo]
PDF set~\cite{Ball:2014uwa}, interfaced with \PYTHIA[8.244] for parton
showering and hadronisation using the A14 set of tuned parameters and
\NNPDF[2.3lo] PDF set.
Following the same methodology as Ref.~\cite{HDBS-2019-29}, signal
templates with coupling modifiers ($\kl \neq 1$, $\kVV \neq 1$) are
obtained by linear combination of six samples with different values
for the \kl\ and \kVV\ parameters.  For the
reweighted VBF signal points, the N$^3$LO to LO cross-section ratio at
the SM value is calculated, and this factor is applied to the
cross-sections at each \kl\ and \kVV\ point.  To assess parton
showering uncertainties, alternative LO samples are generated using
\MADGRAPH[2.7.3] with the \NNPDF[3.0nlo] PDF set, interfaced to
\HERWIG[7.2.1] with the \HERWIG[7.1] default set of tuned parameters
and \MMHT[lo] PDF set for parton showering and hadronisation.
 
The dominant background process to the ML channels is diboson ($VV$)
production, where $V$ refers to production of an electroweak boson
($W$ or $Z/\gamma^*$).  This background is estimated from simulation and
normalised to data in control regions, as described
in Section~\ref{sec:backgrounds}.  Background processes involving
non-prompt leptons, leptons with a wrongly assigned charge, or
misidentified hadronic taus are also important backgrounds to these
channels, and are estimated by using data-driven methods.  The \bbllll\ channel also has
substantial contributions from top quark pair production (\ttbar),
including in association with a $Z$~boson (\ttZ), and $Z$~boson
production in association with jets (\Zjet).  Non-resonant
\yy\ production is the dominant background in the \yyML\ channels,
where the components of this `\yy-continuum' background are
\yy\ production in association with a vector boson (\Vyy), a top quark
pair (\ttyy), or jets (\yyjets).  MC simulations of these three
processes are used when training BDTs to separate signal from
background (as described in Section~\ref{sec:strategy}), while samples
with \yy\ production in association with one or two jets are used to
derive uncertainties in the background estimate, as described in
Section~\ref{subsec:background_yy}.
 
Single Higgs boson production is considered as a background to all
channels, and is significant for the \bbllll\ and \yyML\ channels.
Higgs boson production in association with a vector boson ($VH$)
process is the dominant source of single Higgs backgrounds in most
channels and contributes between 70\% and 90\% of the total single
Higgs background in all channels except the \yytau\ channel where it
is approximately 50\%, and the \bbllll\ channel where it is
negligible.  Gluon--gluon fusion production is negligible in all
channels except in the \yytau\ channel where it contributes approximately
40\% of the total single Higgs boson background, and in the
\bbllll\ channel where it is the dominant source of single Higgs boson
production, contributing around 50\% of the total single Higgs
background.  Higgs boson production in association with a top quark
pair (\ttH) contributes between 7\% and 30\% of the total single Higgs
background in all channels.
 
Simulated samples are produced for the different signal and background
processes using the configurations shown in Table~\ref{tab:mcconfig}.
Details of the samples used to estimate the systematic
uncertainties associated with the generators are shown in parentheses.
All samples include leading-logarithm photon emission, either modelled
by the parton shower generator or by \PHOTOS~\cite{Golonka:2005pn}.
The \SHERPA[2.2.4]~\cite{Bothmann:2019yzt} diphoton plus jets (\yyjets) sample is
simulated with NLO-accurate matrix elements for up to one parton, and
LO-accurate matrix elements for up to three partons are calculated
with the \textsc{Comix}~\cite{Gleisberg:2008fv} and
\OPENLOOPS~\cite{Buccioni:2019sur,Cascioli:2011va,Denner:2016kdg}
libraries.  Both the \yyjets\ and diphoton plus vector boson (\Vyy)
samples are matched with the \SHERPA parton
shower~\cite{Schumann:2007mg} using the \MEPSatNLO
prescription~\cite{Hoeche:2011fd,Hoeche:2012yf,Catani:2001cc,Hoeche:2009rj}
with a dynamic merging cut~\cite{Siegert:2016bre} of \qty{10}{\GeV}.
Photons are required to be isolated according to a smooth-cone
isolation criterion~\cite{Frixione:1998jh}. These samples are generated
using the \NNPDF[3.0nnlo] PDF set~\cite{Ball:2014uwa}, along with the
dedicated set of tuned parton-shower parameters developed by the
\SHERPA authors.

\begin{DIFnomarkup}
\begin{table}[!htp]
\begin{center}
\caption{
The configurations used for event generation of signal and
background processes.  The samples used to estimate the
systematic uncertainties are indicated in grey and enclosed
in parentheses.  The matrix element (ME) order
refers to the order in the strong coupling constant of the
perturbative calculation.  Tune refers to the
set of tuned parameters used by the parton shower generator.
\MGfive\ refers to \MGNLO~\cite{Alwall:2014hca}.  \MePs\ is the
method used in \SHERPA\ to match the matrix element to the
parton shower.  Samples using \PYTHIA[8] have heavy flavour
hadron decays modelled by
\EVTGEN[1.2.0]~\cite{Lange:2001uf}.
}
{\small
\setlength\tabcolsep{1.5pt}
\begin{tabular}{llllll} \toprule
Process                  & Generator                                 & ME order     & Parton shower           & PDF                              & Tune                     \\ \toprule
\textbf{Signal}          & & & & & \\
\ggF~\HH                 & \POWHEGBOX[v2]                            & NLO          & \PYTHIA[8]              & PDF4LHC15\textsc{NLO}            & A14                      \\
& \grey{(\POWHEGBOX[v2])}                   & \grey{(NLO)} & \grey{(\HERWIG[7])}     & \grey{(\MMHT[lo])}               & \grey{(\HERWIG[7] default)} \\
\VBF~\HH                 & \MGfive                                   & LO           & \PYTHIA[8]              & \NNPDF[3.0nlo]                   & A14                      \\
& \grey{(\MGfive)}                          & \grey{(LO)}  & \grey{(\HERWIG[7])}     & \grey{(\MMHT[lo])}               & \grey{(\HERWIG[7] default)} \\ \midrule
\textbf{Top quark}       & & & & & \\
$\ttbar$                 & \POWHEGBOX[v2]~\cite{Frixione:2007nw,Nason:2004rx,Frixione:2007vw}
& NLO          & \PYTHIA[8]              & \NNPDF[3.0nlo]                   & A14                      \\
& \grey{(\POWHEGBOX[v2])}                   & \grey{(NLO)} & \grey{(\HERWIG[7])}     &\grey{(\NNPDF[3.0nlo])}           & \grey{(\HERWIG[7] default)} \\
$\ttbar t$               & \MGfive                                   & LO           & \PYTHIA[8]              & \NNPDF[2.3lo]                    & A14                      \\
$\ttbar \ttbar$          & \MGfive                                   & NLO          & \PYTHIA[8]              & \NNPDF[3.1nlo]                   & A14                      \\
& \grey{(\SHERPA[2.2.10])}                  & \grey{(NLO)} & \grey{(\SHERPA)}        &\grey{(\NNPDF[3.0nnlo])}          & \grey{(\SHERPA default)} \\
Single top               & \POWHEGBOX[v2]~\cite{Re:2010bp,Alioli:2009je} & NLO      & \PYTHIA[8]              & \NNPDF[3.0nlo]                   & A14                      \\
($t$-, $Wt$, $s$-channel)&                                           &              &                         &                                  &                          \\
$\ttbar WW$              & \MGfive                                   & LO           & \PYTHIA[8]              & \NNPDF[2.3lo]                    & A14                      \\
$\ttbar W$               & \SHERPA[2.2.10]                           & NLO          & \SHERPA                 & \NNPDF[3.0nnlo]                  & \SHERPA default          \\
& \grey{(\MGfive)}                          & \grey{(NLO)} & \grey{(\PYTHIA[8])}     &\grey{(\NNPDF[3.0nlo])}           & \grey{(A14)}             \\
$tW, t\Zy$               & \MGfive                                   & NLO          & \PYTHIA[8]              & \NNPDF[2.3lo]                    & A14                      \\
$\ttbar\Zy (\Zll)$       & \MGfive                                   & NLO          & \PYTHIA[8]              & \NNPDF[3.0nlo]                   & A14                      \\ \midrule
\textbf{Vector boson}    & & & & & \\
\Wj,\Zjet                & \SHERPA[2.2.1]                            & NLO          & \SHERPA                 & \NNPDF[3.0nlo]                   & \SHERPA default          \\
\Zll                     & \SHERPA[2.2.1]                            & NLO          & \SHERPA                 & \NNPDF[3.0nlo]                   & \SHERPA default          \\
$VV$, $qqVV$,$VVV$       & \SHERPA[2.2.2]                            & NLO          & \SHERPA                 & \NNPDF[3.0nnlo]                  & \SHERPA default          \\ \midrule
\textbf{Photon}          & & & & & \\
\yyjets                  & \SHERPA[2.2.4]                            & NLO          & \MePs                   & \NNPDF[3.0nnlo]                  & \SHERPA dedicated        \\
\Vyy                     & \SHERPA[2.2.4]                            & LO           & \MePs                   & \NNPDF[3.0nnlo]                  & \SHERPA dedicated        \\
\ttyy                    & \MGfive                                   & LO           & \PYTHIA[8]              & \NNPDF[2.3lo]                    & A14                      \\ \midrule
\textbf{Single Higgs boson}    & & & & & \\
\ggF~$H$                 & \POWHEG                                   & NLO          & \PYTHIA[8]              & PDF4LHC15 NNLO                   & AZNLO~\cite{STDM-2012-23}\\
\VBF~$H$                 & \POWHEG                                   & NLO          & \PYTHIA[8]              & PDF4LHC15 NNLO                   & AZNLO                    \\
\VH\ (inclusive)         & \POWHEGBOX[v2]                            & NLO          & \PYTHIA[8]              & \NNPDF[3.0nlo]                   & A14                      \\
\VH\ (\Hyy)              & \POWHEG                                   & NLO          & \PYTHIA[8]              & PDF4LHC15 NNLO                   & AZNLO                    \\
\ttH                     & \POWHEGBOX[v2]                            & NLO          & \PYTHIA[8]              & \NNPDF[3.0nlo]                   & A14                      \\
& \grey{(\POWHEGBOX[v2])}                   & \grey{(NLO)} & \grey{(\HERWIG[7])}     &\grey{(\NNPDF[3.0nlo])}           & \grey{(H7UE-MMHT~\cite{Bellm:2015jjp})}\\
$bbH$                    & \POWHEGBOX[v2]                            & NLO          & \PYTHIA[8]              & \NNPDF[3.0nlo]                   & A14                      \\
$tHb$+jet(s)             & \MGfive                                   & NLO          & \PYTHIA[8]              & \NNPDF[3.0nlo]                   & A14                      \\
$tHW$                    & \MGfive                                   & NLO          & \PYTHIA[8]              & \NNPDF[3.0nlo]                   & A14                      \\  \bottomrule
\end{tabular}
}
\label{tab:mcconfig}
\end{center}
\end{table}
\end{DIFnomarkup}


\section{Object definitions}
\label{sec:objects}

 
Vertices from \pp interactions are
reconstructed~\cite{ATL-PHYS-PUB-2015-026} using at least two inner detector
tracks with ${\pt > 500\,\MeV}$.  In the ML channel analyses, the hard
scatter primary vertex is defined to be the vertex with the largest
sum of squared track momenta, $\sum\pt^{2}$.  For the \yyML\ channel
analyses, the hard scatter primary vertex is chosen using a neural
network that uses information about inner detector tracks and the diphoton
system~\cite{HIGG-2013-08}.
 
Electrons, muons, hadronic taus, photons, jets (including those
containing $b$-hadrons) and missing transverse energy, \met, are used
in this search.  Their reconstruction and identification are described
below.  Three selection definitions are used for both electrons and
muons in the ML channel analyses -- `Baseline', `Loose', and
`Tight' -- that are optimised for use in different channels and
regions.  A fourth definition is used in the \yyML\ channels.
Their definitions are summarised in Table~\ref{tab:lepton_defs}.
 
Electrons are reconstructed and identified by matching inner detector
tracks to energy deposits measured in the electromagnetic
calorimeter~\cite{EGAM-2018-01}.  Electron candidates are
required to have $\pt > 10$\,\GeV and $|\eta| < 2.47$, excluding the
calorimeter transition region $1.37 < |\eta| < 1.52$.  The
minimum \pt\ requirement is lowered to 4.5\,\GeV in the \bbllll\
channel, where one of the $Z$\ bosons is produced off-shell and as
such its decay products are typically produced with low \pt.  Electron
candidates are identified using a likelihood technique and both
Baseline and Loose candidates are required to satisfy a loose
identification working point, which, in combination with additional
track hit requirements applied to ensure that the track is high
quality, provides an overall electron selection efficiency of 93\% in
a \Zee\ sample~\cite{EGAM-2021-01}.  The Tight electrons are required
to satisfy a tight identification working point~\cite{EGAM-2021-01}
that is 80\% efficient at selecting electrons in \Zee\ events.  No
isolation requirements are applied as part of the Baseline definition,
but Loose (Tight) electrons are required to satisfy loose (tight)
isolation working points of a `Prompt Lepton Veto' (PLV) BDT designed
to reject non-prompt electrons~\cite{HIGG-2017-02}.  Several signal
regions are defined based on the relative charge of two leptons, so a
charge ($Q$) mis-ID BDT is used to reject electron candidates where
the charge is likely to have been wrongly attributed. The chosen
working point of the charge mis-ID BDT
~\cite{HIGG-2017-02,EGAM-2018-01}\ provides an electron charge mis-ID
probability of typically less than 0.2\% for a 95\% signal efficiency
in \Zee\ events.  Contributions from converted photons are rejected
using an ambiguity solving algorithm based on track
information~\cite{HIGG-2017-02,EGAM-2018-01}.  A fourth electron
definition working point is used in the \yyML\ channels, where the
electron candidates are required to have $\pt>10$\,\GeV, and satisfy
the medium working point of the likelihood based
identification~\cite{EGAM-2021-01} that is 88\% efficient
at selecting electrons in \Zee\ events.  Isolation requirements are
applied, based on the presence of tracks in a cone of \pT-dependent
size around the electron and of calorimetric energy deposits in a
fixed-size cone~\cite{EGAM-2021-01}.  The isolation requirements are
approximately 85\% efficient for electrons with \ET\ of 10\,\GeV, and
fully efficient for electrons with $\ET > 40$\,\GeV.
 
Muon candidates are reconstructed from tracks in the MS, which are
matched to inner detector tracks where available.  Baseline muon
candidates are required to have $\pt > 10$\,\GeV and $|\eta| < 2.5$.
The minimum \pt\ requirement is lowered to 3\,\GeV in the \bbllll\
channel to increase acceptance of muons from the decay of an off-shell
$Z$\ boson.  Baseline and Loose muons are required to satisfy a loose
identification working point that is typically 98\% efficient at
selecting prompt muons.  The identification is tightened to a medium
working point that is around 97\% efficient for the Tight muon
definition~\cite{MUON-2018-03}.  Similarly to electrons, Loose and
Tight muons are required to satisfy correspondingly strict working
points of the PLV~\cite{MUON-2018-03}.  The loose (tight) PLV working
points are 81\% (57\%) efficient at selecting the lowest \pt\ prompt
muons, rising to 93\% (87\%) for muons with $\pt > 20$\,\GeV.  Muons
used in the \yyML\ channels are required to have $\pt>10$\,\GeV,
satisfy the same medium identification working point as is used in the
ML channels, and loose isolation requirements that are based on the
presence of particle-flow objects~\cite{PERF-2015-09} in a cone
of \pT-dependent size around the muon.  The isolation requirements are
95--99\% efficient at selecting prompt muons in the \pT\ regions used
in the analysis.
 
To further reduce contributions from non-prompt electrons and muons,
cuts on the transverse and longitudinal impact parameters with respect
to the primary vertex, $|d_{0}|$\ and $|z_{0}|$ respectively, are
applied to all candidates.  Electrons (muons) are required to have
$|d_{0}|/\sigma_{d_{0}} < 5 (3)$\ and $|z_{0} \sin \theta| < 0.5$~mm
(where $\sigma_{d_{0}}$\ is the uncertainty on the reconstructed
$d_{0}$, and $\theta$\ is the polar angle of the track).
 
\begin{DIFnomarkup}
\begin{table}[!t]
\begin{center}
\caption{Electron and muon candidate definitions used in the analysis. A `--' indicates that a cut is not applied, and `N/A' indicates that a requirement is not applicable.}
{\small
\begin{tabular}{l|c|c|c|c|c|c|c|c} \toprule
\multirow{3}{*}{}              & \multicolumn{4}{c|}{Electrons}                                         & \multicolumn{4}{c}{Muons}                                             \\
& Baseline                    & Loose      & Tight          & \yyML      & Baseline                   & Loose       & Tight     & \yyML          \\
& (B)                         & (L)        & (T)            & (P)        & (B)                        & (L)         & (T)       & (P)            \\ \toprule
Minimum \pt                    & \multicolumn{4}{c|}{10\,\GeV}                                          & \multicolumn{4}{c}{10\,\GeV}                                          \\
& \multicolumn{4}{c|}{\scriptsize{(\bbllll\ channel: 4.5\,\GeV)} }       & \multicolumn{4}{c}{\scriptsize{( \bbllll\ channel: 3\,\GeV)}}         \\
$\eta$                         & \multicolumn{4}{c|}{$|\eta|<1.37~\text{or}~1.52<|\eta|<2.47$}          & \multicolumn{4}{c}{$|\eta|<2.5$}                                      \\
Identification                 & \multicolumn{2}{c|}{Loose}               & Tight          & Medium     & \multicolumn{2}{c|}{Loose}               & \multicolumn{2}{c}{Medium} \\
Isolation                      & --                          & PLV loose  & PLV tight      & Loose      & --                         & PLV loose   & PLV tight & Loose          \\
$Q$\ mis-ID BDT                & --                          & \multicolumn{2}{c|}{\cmark} & --         & \multicolumn{4}{c}{N/A}                                               \\
$e/\gamma$~ambiguity           & --                          & \multicolumn{2}{c|}{\cmark} & --         & \multicolumn{4}{c}{N/A}                                               \\
$|d_0|/\sigma_{d_0}$            & \multicolumn{4}{c|}{$<5$}                                              & \multicolumn{4}{c}{$<3$ }                                             \\
$|z_0 \sin \theta|$            & \multicolumn{4}{c|}{$<0.5$~mm}                                         & \multicolumn{4}{c}{$<0.5$~mm}                                         \\ \bottomrule
\end{tabular}
}
\label{tab:lepton_defs}
\end{center}
\end{table}
\end{DIFnomarkup}
 
The visible hadronic tau decay (\tauhadvis) reconstruction
algorithm~\cite{PERF-2013-06} is seeded from jets formed using the
\antikt\ algorithm~\cite{Fastjet,Cacciari:2008gp} with a radius
parameter $R = 0.4$, and clusters of calorimeter cells calibrated using
a local hadronic calibration (LC)~\cite{ATL-PHYS-PUB-2009-001-2} as
inputs.  The \tauhvis\ candidates are required to have
$\pt > 20$\,\GeV and $|\eta| < 2.5$.  The calorimeter transition region
($1.37 < |\eta| < 1.52$) is vetoed. A set of BDTs are used to determine
whether tracks in a cone with radius $R = 0.4$ around the \tauhvis\
axis are consistent with coming from a hadronic decay of a tau.
Selected \tauhvis\ candidates are required to have either one or three
associated tracks (or `prongs'), with a total charge of $\pm 1$.
Recurrent neural networks (RNNs) are used to identify
\tauhvis\ candidates and reject backgrounds~\cite{ATL-PHYS-PUB-2019-033}.
A loose identification working point is used in the \yyML\ channels,
providing an efficiency of 85\% (75\%) for one-prong
(three-prong) \tauhvis.  In the ML channels, the medium working point
is used that has an efficiency of 75\% (60\%) for one-prong
(three-prong) \tauhvis.  A separate BDT is used to reject electrons
that are misidentified as one-prong
\tauhvis\ candidates, with an efficiency of about 95\% for real
hadronic taus~\cite{ATLAS-CONF-2017-029}.
 
Anti-ID \tauhvis\ objects, defined as a \tauhvis\ with modified
identification requirements, are used to estimate the backgrounds from
jets misidentified as \tauhvis\ in the \lltautau\ and \ltautau\
channels, as described in Section~\ref{subsec:background_taus}.
Anti-ID \tauhvis\ objects are reconstructed, and their
energy is calibrated in the same way as
\tauhvis\ candidates, and they must satisfy the nominal
\tauhvis\ kinematic and track selection criteria. They are
required to satisfy a very loose RNN identification requirement,
corresponding to an efficiency of approximately 99\% for
\tauhvis~\cite{ATL-PHYS-PUB-2019-033}, but fail to satisfy the
nominal RNN requirement applied to the \tauhvis\ candidates.
 
Photon candidates are required to have $\et>25$\,\GeV and
$|\eta|<2.37$.  Photons inside the transition region of the calorimeter
($1.37 < |\eta| <1.52$) are rejected.  Photon identification is based
on the lateral shower profile of the energy deposits in the first and
second electromagnetic calorimeter layers and on the energy leakage
fraction in the hadronic calorimeter.  The photon candidates are also
required to satisfy a tight working point of this identification
algorithm, which is tuned for converted and unconverted photons
separately~\cite{EGAM-2018-01}.  Loose isolation requirements, based
on calorimeter energy clusters and tracks in a cone with radius $R = 0.2$\
around the photon are also applied~\cite{EGAM-2018-01}.  For isolated
photons with \pt\ between 30\,\GeV and 250\,\GeV, the identification
efficiency for unconverted and converted photons ranges from 84\% to
98\%, when evaluated on a sample of \Zlly\ events~\cite{EGAM-2018-01}.
 
Reconstructed jets are based on particle-flow objects built from
noise-suppressed positive-energy topological clusters in the
calorimeter and reconstructed tracks~\cite{PERF-2015-09}.  The
\antikt\ algorithm with a radius parameter of $R=0.4$ is used.
Jets are required to have $|\eta|<2.5$\ (extended to $|\eta|<4.4$\
for channels with photons) and $\pt > 25$\,\GeV.  To further suppress
jets produced in concurrent $pp$\ interactions, each jet within the
tracking acceptance of $|\eta|<2.4$, and with $\pT < 60$\,\GeV, is
required to satisfy the tight jet-vertex tagger~\cite{PERF-2014-03}
criteria used to identify the jet as originating from the selected
primary vertex of the event.
 
Jets containing $b$-hadrons, $b$-jets, are identified using a
deep-learning neural network, DL1r~\cite{FTAG-2019-07} that combines
information about the impact parameters of inner detector tracks, the presence of
displaced secondary vertices, and reconstructed flight-paths of $b$-
and $c$-hadrons inside the jet.  Jets with $|\eta|<2.5$ are
considered for $b$-tagging.  A working point that gives 77\%
efficiency to identify jets associated with a $b$-hadron in
simulated \ttbar\ events is used to select, or veto, $b$-jets.  At
this working point, the light-jet (charm-jet) rejection measured
in \ttbar\ simulation is about a factor of 130
(4.9)~\cite{FTAG-2018-01}.  The DL1r algorithm is calibrated using a
likelihood-based method for each jet type~\cite{FTAG-2018-01}, and
correction factors are applied to the simulated event samples to
compensate for differences between data and simulation in the
$b$-tagging efficiency for $b$-, $c$- and light-flavour jets.  The
energy of $b$-tagged jets containing a muon is corrected to account
for the fact that a muon typically only deposits a small fraction
of its energy in the calorimeters.  In addition, the
undetected energy of the neutrinos and out-of-cone effects are
corrected for with scale factors derived as a function of the
$b$-jet \pT\ from a \ttbar\ MC sample. The two corrections together
improve the resolution of the invariant mass of the two jets with the
highest $b$-tagging score (\mbb) by about 18\% for SM \HH\
signal events that include a $H\rightarrow b\bar{b}$\ decay. The
procedure closely follows the one used in Ref.~\cite{HIGG-2016-29}.
 
An overlap-removal procedure is applied to resolve
ambiguities between independently reconstructed electrons, muons,
(anti-ID) \tauhvis, photons and jets.  Any electron
found to share a track with a muon is removed, as is any
(anti-ID) \tauhadvis\ within $\dR = 0.2$\ of an electron or muon.  Jets
found within $\dR = 0.2$\ of an electon or (anti-ID) \tauhadvis\ are
removed, and any jet with less than three tracks associated with it is
removed if it is found to be within $\dR = 0.2$\ of a muon.  Any
electron or muon found within $\dR= 0.4$\ of surviving jets is
removed.  Photons are removed if they are found within $\dR = 0.4$\ of
an electron or muon.  Jets found within $\dR = 0.4$\ of a photon are
removed.  All requirements are applied sequentially.  A similar
procedure is applied for the \yyML\ channels, but prioritising
photons.  All electrons and muons that satisfy the `Baseline'
definition are considered as inputs to the overlap-removal procedure
in the ML channels while the \yyML\ channels use the definitions
described in the text and referred to as type `P' in
Table~\ref{tab:lepton_defs}.  The differences between the overlap-removal
procedures means that the ML and \yyML\ channels are not strictly
orthogonal, but no signal or data events are found to satisfy the
selection requirements of more than one channel.
 
The missing transverse energy \met\ in an event is calculated as the
magnitude of the negative vectorial sum of the transverse momenta of
all selected and calibrated physics objects that can be matched to the
primary vertex, after the overlap removal procedure is applied. A
component called the `soft term' is calculated from the residual
tracks that originate from the primary vertex but are not associated
with any other object and is included in the
calculation~\cite{JETM-2020-03}.


\section{Event categorisation and preselection}
\label{sec:events}

 
Events in the ML channels that have final states containing two or
more light leptons are selected by requiring that they satisfy single
lepton or dilepton triggers~\cite{TRIG-2018-05,TRIG-2018-01}.  Events
in the \ltautau\ channel are selected using only the single lepton
triggers.  The single electron (muon) triggers have \pt\ thresholds of
24--26\,\GeV (20--26\,\GeV), depending on the data-taking conditions.
The dilepton triggers require either two electrons, two muons, or one
electron and one muon, and have \pt\ thresholds as low as 12\,\GeV
(8\,\GeV) for the leading (subleading) lepton.  Diphoton
triggers~\cite{TRIG-2018-05} where the leading (subleading) photon is
required to have $\ET > 35$\,\GeV~(25\,\GeV) are used in the \yyML\
channels.  The diphoton triggers used in 2015 and 2016 required that
both the photons satisfy the loose photon identification critera, and
this was tightened to a medium identification working point during
2017--2018 data taking in response to the increased $pp$ interaction
rate.  In all channels, the electrons, muons and photons that fired
the trigger are required to be geometrically matched to corresponding
offline objects.  Electrons and muons that are geometrically matched
to the trigger objects are required to have an offline \pt\ 1\,\GeV\
higher than the threshold of the corresponding trigger.

Events are categorised into sub-channels according to the number of
photons, \tauhvis, and light leptons satisfying the
definitions in Table~\ref{tab:lepton_defs}, after applying the
overlap removal procedure.  The requirements for the
different sub-channels are summarised for the ML channels in
Table~\ref{tab:sr_defs_ml} and for the \yyML\ channels in
Table~\ref{tab:sr_defs_yyml}.  These sets of requirements define the
signal preselection regions that are used for further multivariate
analysis selections used to refine the extraction of signal as
described in Section~\ref{sec:strategy}.  The requirements also form
the basis from which control and validation regions are defined in
order to estimate background contributions, as described in
Section~\ref{sec:backgrounds}.  The contributions of the different
decay modes of the Higgs boson pair to different signal regions after
applying the preselection requirements is shown for the ML channels in
Figure~\ref{fig:channel_flow_ml} and for the \yyML\ channels in
Figure~\ref{fig:channel_flow_yyml}.
 
\begin{DIFnomarkup}
\begin{table}[!tbhp]
\begin{center}
\caption{Selection criteria applied to each ML channel to form the
signal preselection regions. The notation `N$\ell$(X)' refers to the
multiplicity, N, of the different types of lepton (X = B,L,T) as
defined in Table~\ref{tab:lepton_defs}.  The multiplicity
of \tauhvis, jets, and $b$-jets are denoted \Ntau, \Njet,
and \Nbjet, respectively.  When no \pt\ (or \ET) threshold is
specified, the default requirements for each object are used, as
described in Section~\ref{sec:objects}.  Objects are ordered by
decreasing \pt and their index denoted by a subscript.  In
the \bbllll\ channel, the same-flavour, opposite charge (SFOC)
lepton pair with an invariant mass closest to the $Z$\ boson mass is
defined as the lepton pair coming from the on-shell $Z$\ boson decay
(on-shell-$\ell\ell$) while the remaining SFOC lepton pair is
defined as coming from the off-shell $Z$\ decay
(off-shell-$\ell\ell$).  In the \lll\ channel, the lepton with
opposite charge relative to the other two is denoted by
$\ell_{\text{OC}}$.  The same-charge lepton that is nearest to \loc\
in $\Delta R$\ is denoted \lscone\ and the other is \lsctwo.  The
`$Z$-veto' requires that the invariant mass of two SFOC leptons must
satisfy $|m_{\ell\ell}-m_{Z}| > 10$\,\GeV.  An analogous $Z$-veto
requirement is considered for the three-lepton system in the \lll\
channel to remove background processes with
$Z\rightarrow \ell\ell\gamma^{*}(\gamma^{*}\rightarrow\ell^{'}\ell^{'})$\
where one lepton has very low momentum and is not reconstructed.}
{\small
\begin{tabular}{l|c c c c} \toprule
Channel       & $\ell$                                      & \tauhvis                   & Jets                 & $b$-jets                \\ \toprule
 
\bbllll       & 4$\ell$(B)                                  & $\Ntau=0$                  & $\Njet \ge 2$         & $1 \le \Nbjet \le 3$   \\
& $\pt(\ell_1) > 20$\,\GeV                    &                            &                       &                        \\
& $\pt(\ell_2) > 15$\,\GeV                    &                            &                       &                        \\
& $\pt(\ell_3) > 10$\,\GeV                    &                            &                       &                        \\
& $\ell_3$\ or $\ell_4$\ pass loose PLV       &                            &                       &                        \\
& 2 SFOC pairs                                &                            &                       &                        \\
& $50 < \mllonshell < 106$\,\GeV              &                            &                       &                        \\ \\[-1em] 
& $5 < \mlloffshell < 115$\,\GeV              &                            &                       &                        \\ \\[-1em] 
& All 4 pairs $\Delta R(\ell_i,\ell_j)>0.02$  &                            &                       &                        \\
& $\mllllmZ > 10$\,\GeV                       &                            &                       &                        \\ \midrule
 
\lll          & 3$\ell$, sum of charges = $\pm$1            & $\Ntau=0$                  & $\Njet \ge 1$         & $\Nbjet=0$             \\
& \loc(L)                                     &                            &                       &                        \\
& \lscone(T), $\pt>15$\,\GeV                  &                            &                       &                        \\
& \lsctwo(T), $\pt>15$\,\GeV                  &                            &                       &                        \\
& All $\mllSFOS > 12$\,\GeV                   &                            &                       &                        \\
& $Z$-veto                                    &                            &                       &                        \\
& $\mlllmZ > 10$\,\GeV                        &                            &                       &                        \\ \midrule
 
\SSll         & 2$\ell$(T), $\pt>20$\,\GeV, SC              & $\Ntau=0$                   & $\Njet \ge 2$         & $\Nbjet=0$             \\
& $\mll > 12$\,\GeV                           &                             &                       &                        \\ \midrule
\SSlltau      & 2$\ell$(T), $\pt>20$\,\GeV, SC              & $\Ntau=1$                   & $\Njet \ge 2$         & $\Nbjet=0$             \\
& $\mll > 12$\,\GeV                           & $\pt>25$\,\GeV              &                       &                        \\
&                                             & OC to $\ell$                &                       &                        \\ \midrule
 
\lltautau     & 2$\ell$(L), OC                              & $\Ntau=2$, OC               & $\Njet \ge 0$         & $\Nbjet=0$             \\
& $\mll > 12$\,\GeV                           & $\Delta R(\tau_1,\tau_2)<2$ &                       &                        \\
& $Z$-veto                                    &                             &                       &                        \\ \midrule
 
\ltautau      & 1$\ell$(L)                                  & $\Ntau=2$, OC               & $\Njet \ge 2$         & $\Nbjet=0$             \\
&                                             & $\Delta R(\tau_1,\tau_2)<2$ &                       &                        \\ \bottomrule

\end{tabular}
}
\label{tab:sr_defs_ml}
\end{center}
\end{table}
\end{DIFnomarkup}

\begin{DIFnomarkup}
\begin{table}[!tb]
\begin{center} \caption{Selection criteria applied to each \yyML\
channel to form the signal preselection regions. The notation
`$N(\ell\textrm{(P)})$' refers to the multiplicity, $N$, of P-type leptons as
defined in Table~\ref{tab:lepton_defs}.  The multiplicity
of \tauhvis, photons, and $b$-jets are denoted \Ntau, \Ny,
and \Nbjet, respectively.  When no \pt\ or \ET\ threshold is
specified, the default requirements for each object are used, as
described in Section~\ref{sec:objects}.  Photons are ordered in decreasing
\ET and their index denoted by a subscript.  The invariant
masses of dilepton and diphoton systems are denoted $m_{2(\ell,\tau)}$\
and \myy\ respectively.}
{\small
\begin{tabular}{l|c c c c c} \toprule
Channel       & $\ell$              & \tauhvis                   & Photons                      & \met                       & $b$-jets                      \\ \midrule
 
\yyll         & \multicolumn{2}{c}{$N(\ell\textrm{(P)})+\Ntau=2$, OC}  & \multirow{5}{*}{\shortstack[c]{
$\Ny=2$\\
$\ET(\gamma_1)>35$\,\GeV \\
105\,\GeV~$<\myy<$~160\,\GeV\\
$\gamma_1: \pt/\myy>0.35$ \\
$\gamma_2: \pt/\myy>0.25$}}    & $\met>35$\,\GeV       & \multirow{5}{*}{$\Nbjet=0$}   \\
& \multicolumn{2}{c}{$m_{2(\ell,\tau)} > 12$\,\GeV}  &                                    &                       &                               \\ \cmidrule(lr){1-3}\cmidrule(lr){5-5}
 
\yyl          & $N(\ell(P))=1$      & $\Ntau=0$                  &                                    & \yye: $\met>35$\,\GeV &                               \\
&                     &                            &                                    & \yymu: --             &                               \\ \cmidrule(lr){1-3}\cmidrule(lr){5-5}
\yytau        & $N(\ell(P))=0$      & $\Ntau=1$                  &                                    & $\met>35$\,\GeV       &                               \\ \bottomrule
 
\end{tabular}
}
\label{tab:sr_defs_yyml}
\end{center}
\end{table}
\end{DIFnomarkup}

\begin{figure}[!htb]
\centering \includegraphics[width=1.0\textwidth]{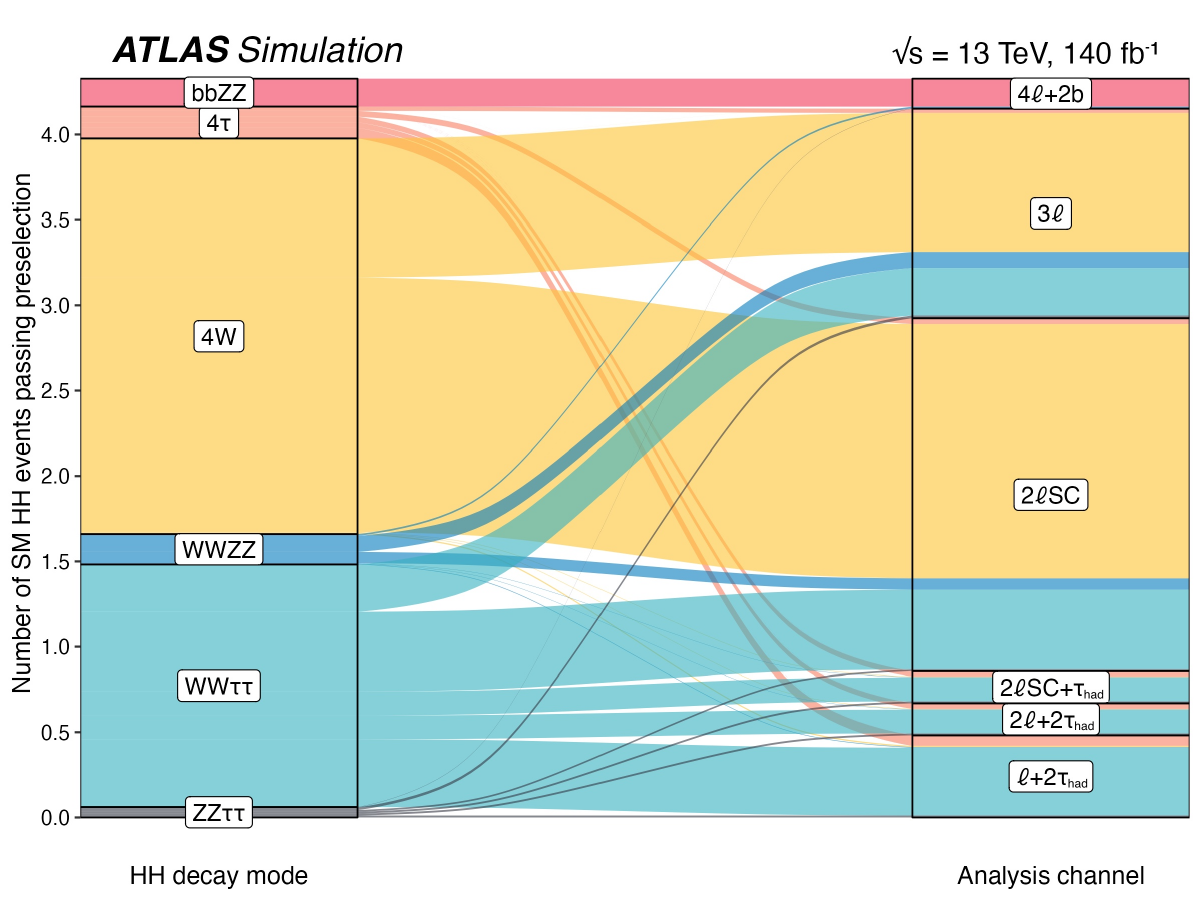}
\caption{Number of \ggF\ and \VBF\ SM \HH\ signal events satisfying
the preselection requirements from the targeted \HH\ decay modes
on the left and their acceptance into the different ML search
channels on the right. The \HHZZZZ\ decay mode contributes less
than 0.1\% of preselected \HH\ events and is not shown. }
\label{fig:channel_flow_ml}
\end{figure}
 
\begin{figure}[!htb]
\centering \includegraphics[width=1.0\textwidth]{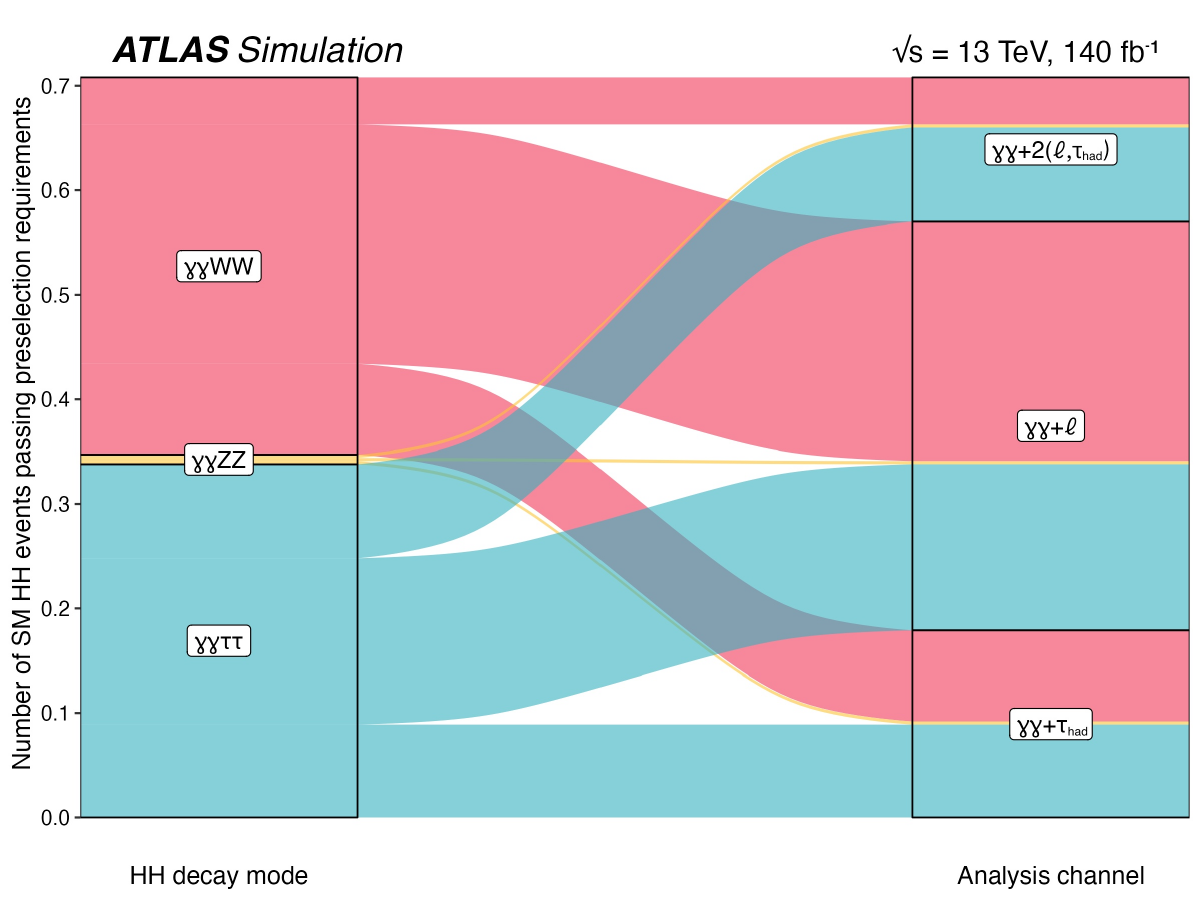}
\caption{Number of \ggF\ and \VBF\ SM \HH\ signal events satisfying
the preselection requirements from the targeted \HH\ decay modes
on the left and their acceptance into the different \yyML\ search
channels on the right. }
\label{fig:channel_flow_yyml}
\end{figure}


\section{Search strategy }
\label{sec:strategy}

All channels except the \yyll\ channel use BDTs with the Gradient
Boost~\cite{gradient_boost} algorithm to separate signal from
background processes.  BDTs are optimised separately for each channel,
in terms of the input variables and hyperparameters, using the area
under the receiver operating characteristic (ROC) curve as the
performance metric.  Information about the kinematics of objects in
the event are used as inputs to the BDT, as well as variables that
probe the kinematic relationships between the objects, for example the
angular separation or invariant mass of two or more objects. The
complete list of all the variables used as inputs to the BDTs in the
different channels is provided in the Appendix.  No BDT or further
event selection is used in the \yyll\ channel analysis due to the low
event yields and the signal region is therefore defined by the
preselection requirements described in Section~\ref{sec:events}.  Both
the ggF~\HH\ and VBF~\HH\ processes are considered as signal in all
channels.  All background processes are included when training the
BDTs used for the \bbllll, \lll, and \yyML\ channel analyses, while
the ML channels with hadronic taus in the final state
(\SSlltau, \lltautau\ and \ltautau) are trained only
against the dominant diboson (\VV) background process.  For the \SSll\
channel, better separation of signal from background was demonstrated
by training three BDTs separately to distinguish the \HH\ signal from
the main \VV, \ttbar, and \Zjet\ background processes.  The three
output BDT score distributions are then used as the input variables to
a fourth BDT that is trained against all background processes and used
as the final discriminating variable to separate signal from
background.  This strategy was found to improve the sensitivity of
the \SSll\ channel over the use of a single-BDT approach.
 
The BDT output score distribution is used as the final discriminant in
each of the ML channels.  The full distribution is used in
the \bbllll, \lltautau, and \ltautau\ channels, while the \lll, \SSll,
and \SSlltau\ channels use the high-BDT-score region as the signal
region, and use the low-BDT-score region to validate the background
model or constrain background processes, as described in
Section~\ref{sec:backgrounds}.  The BDT output score in the signal
region is shown for each channel in Figure~\ref{fig:results_ml}.  In
the \yyl\ and \yytau\ channels the BDT score is used to define `Tight'
($0.6 < $~BDT score~$\le 1$), `Medium' ($0 <$~BDT score~$\leq 0.6$) and
`Loose' ($-1 \le$~BDT~score~$\leq 0$) signal regions, where the BDT score
cuts are optimised to maximise the significance, and the Loose BDT score
regions are used primarily as a background control region in the fit.
The \myy\ distribution is used as the final discriminant in each
\yyML\ signal region, as shown in Figure~\ref{fig:results_yyml}.
The distributions in Figures~\ref{fig:results_ml}\
and~\ref{fig:results_yyml}\ are shown after applying the likelihood
fit to data (i.e. `post-fit') under the background-only hypothesis as described in
Section~\ref{sec:results}.
 
\begin{figure}[!tbhp]
\centering
\subfloat[]{ \includegraphics[width=0.32\textwidth]{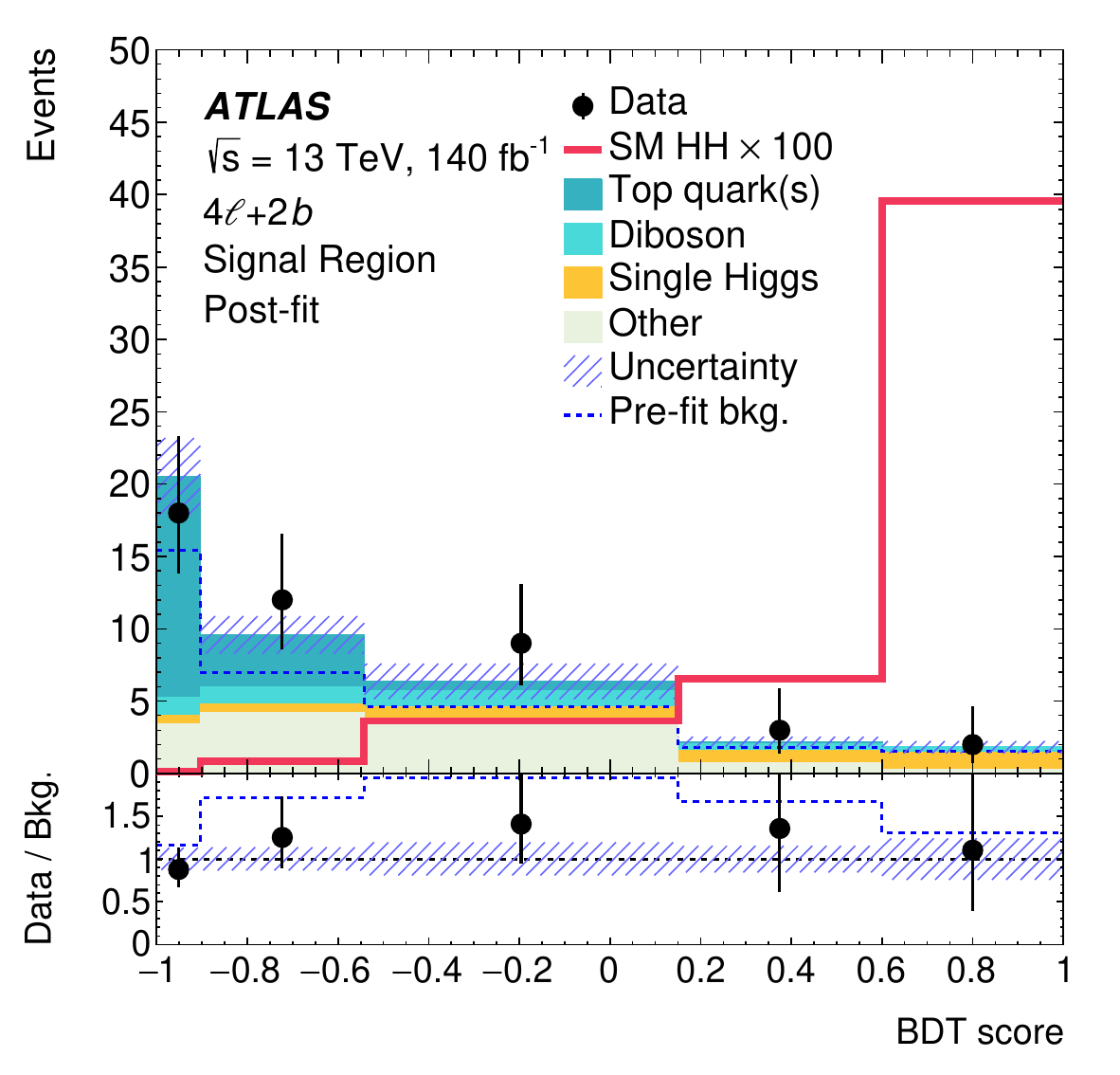}
}
\subfloat[]{ \includegraphics[width=0.32\textwidth]{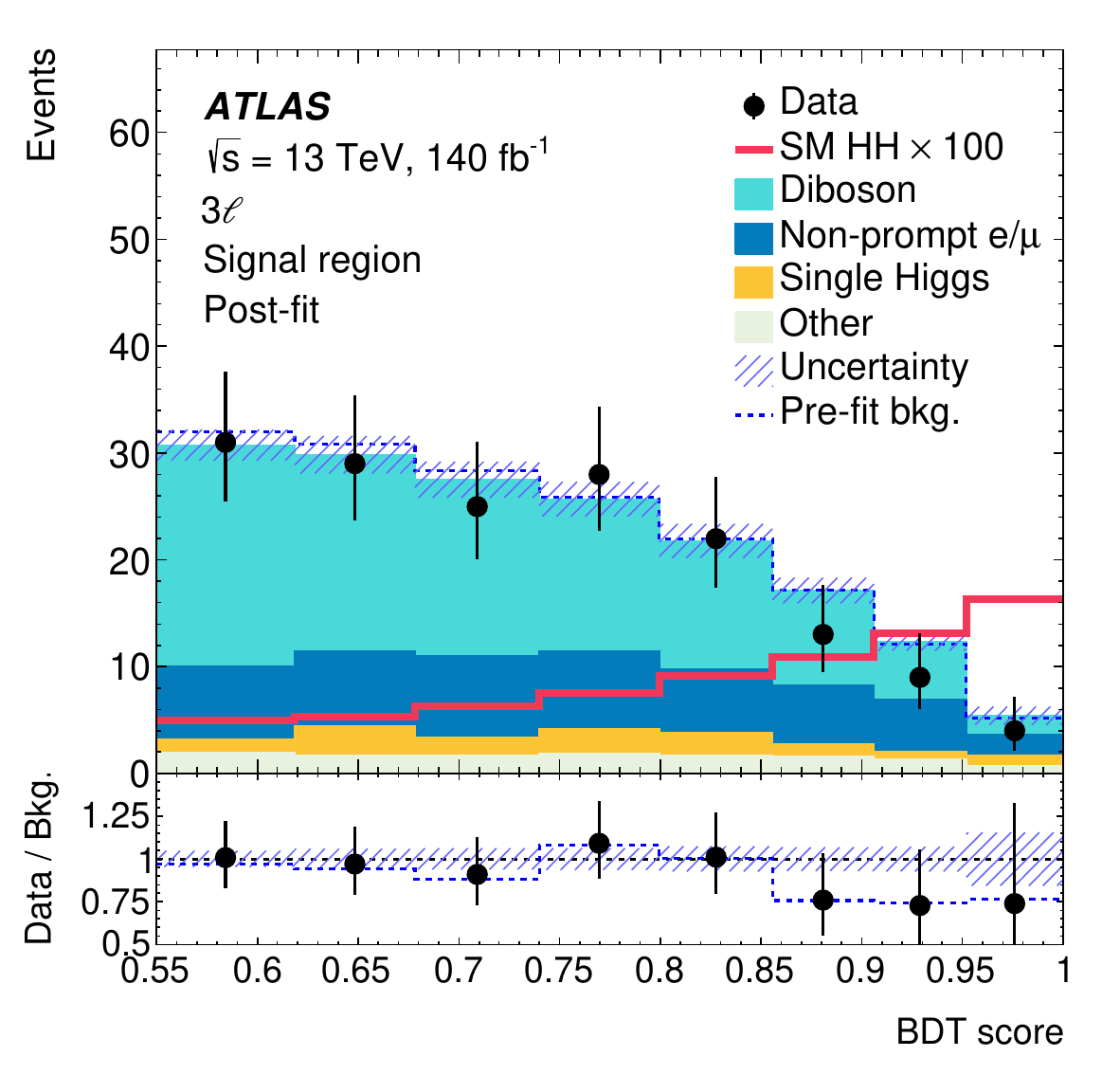}
}
\subfloat[]{ \includegraphics[width=0.32\textwidth]{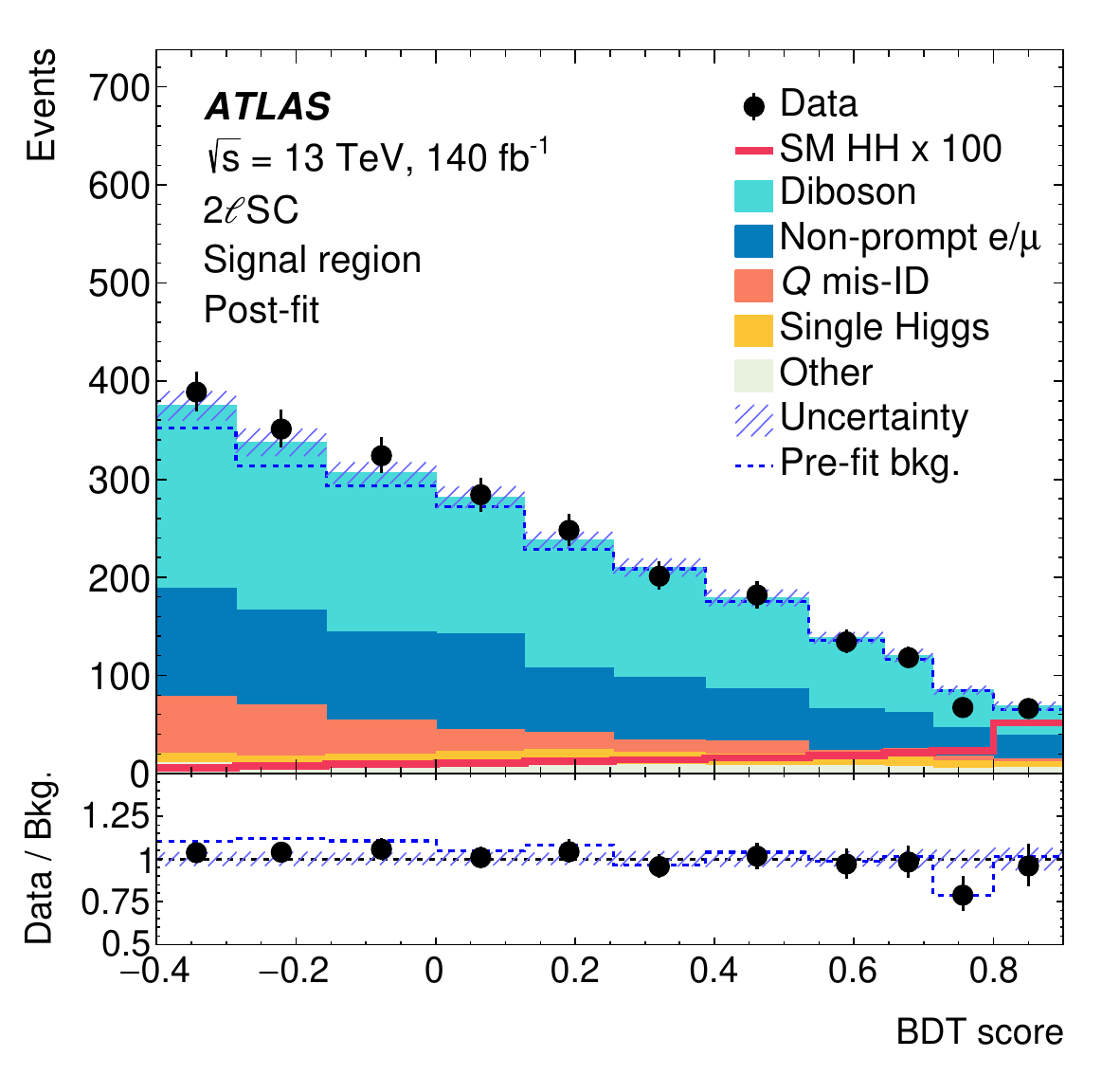}
}\\
\subfloat[]{ \includegraphics[width=0.32\textwidth]{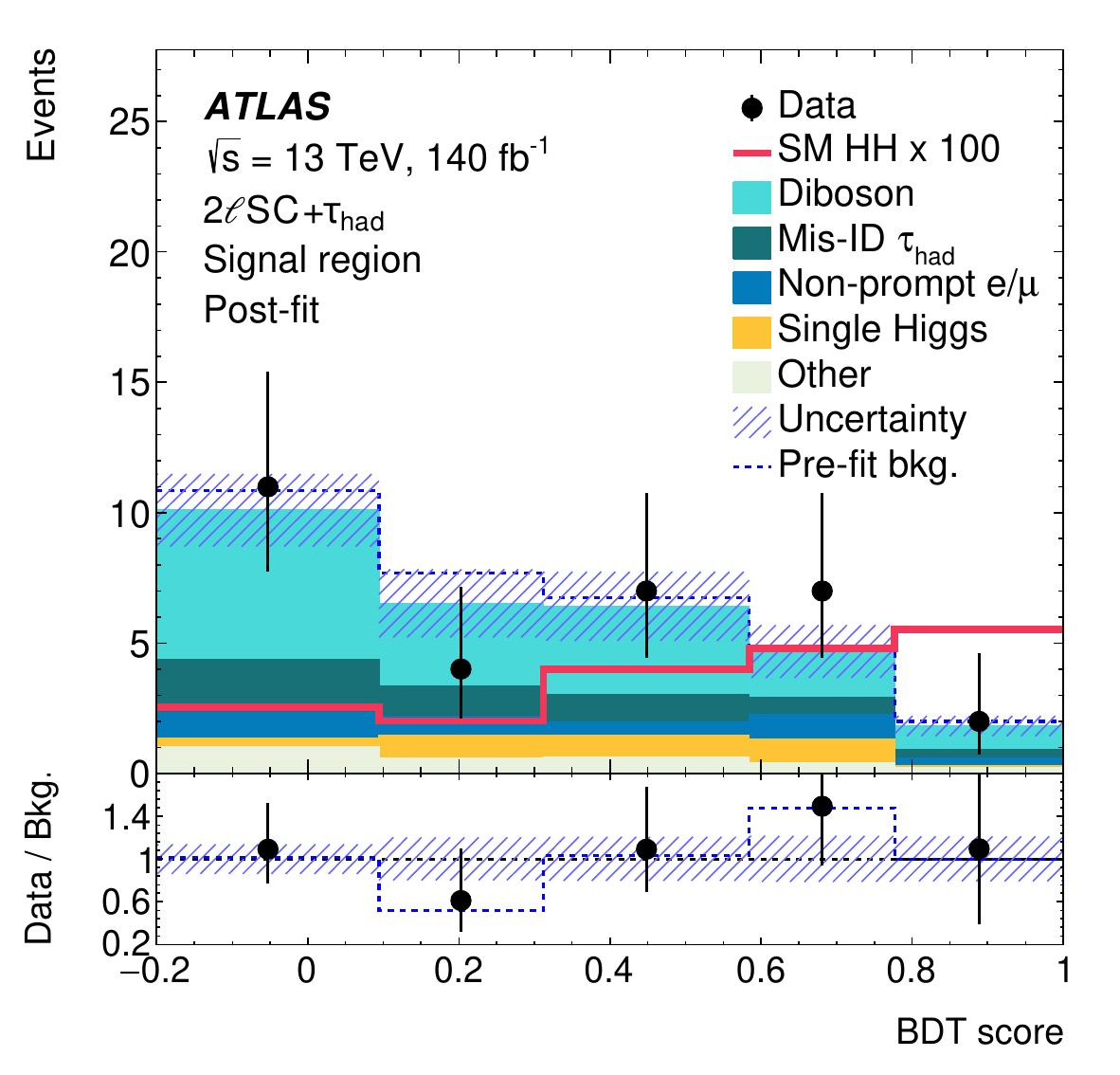}
}
\subfloat[]{ \includegraphics[width=0.32\textwidth]{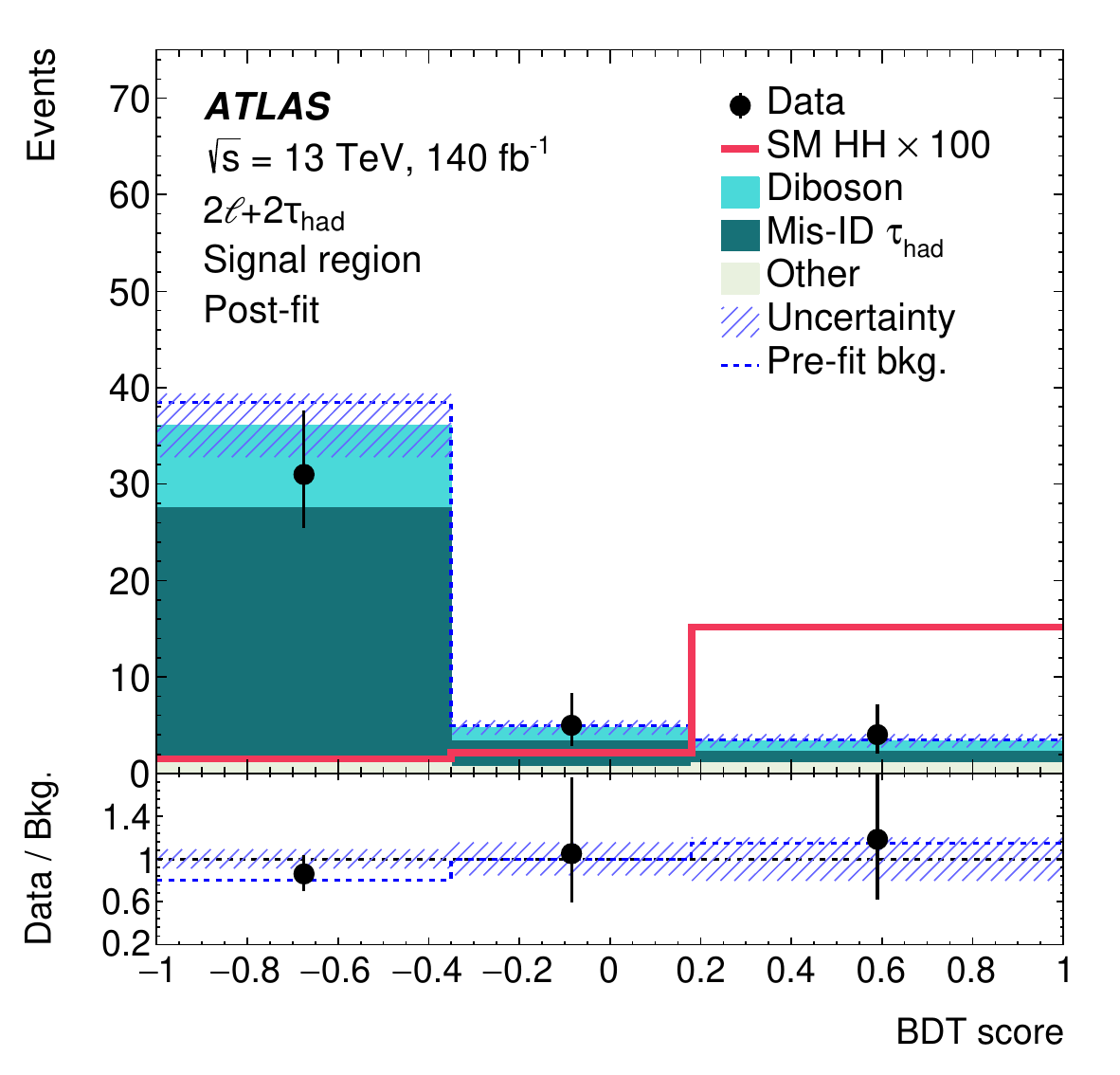}
}
\subfloat[]{ \includegraphics[width=0.32\textwidth]{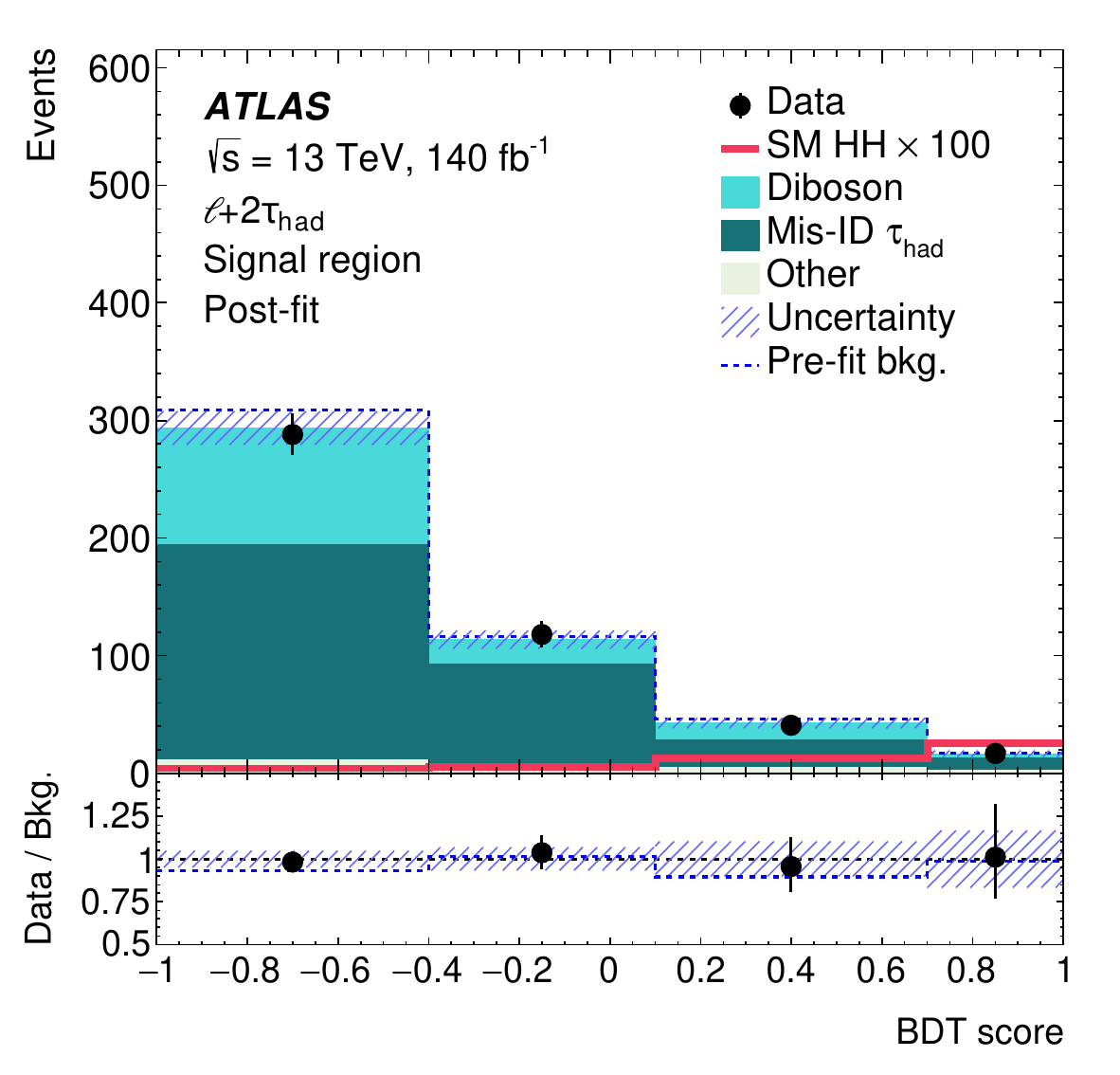}
}
\caption{Distributions of the BDT output score in the signal regions
of the (a)~\bbllll, (b)~\lll, (c)~\SSll, (d)~\SSlltau,
(e)~\lltautau, and (f)~\ltautau\ channels, after applying the
likelihood fit to data under the background-only hypothesis as
described in Section~\ref{sec:results}.  The total pre-fit
background (and its ratio to data) is also shown, as is the SM \HH\
signal scaled up by a factor of 100.  The uncertainty bands include
all sources of statistical and systematic uncertainties in the
background prediction.
}
\label{fig:results_ml}
\end{figure}

\begin{figure}[!tbhp]
\centering
\subfloat[]{ \includegraphics[width=0.32\textwidth]{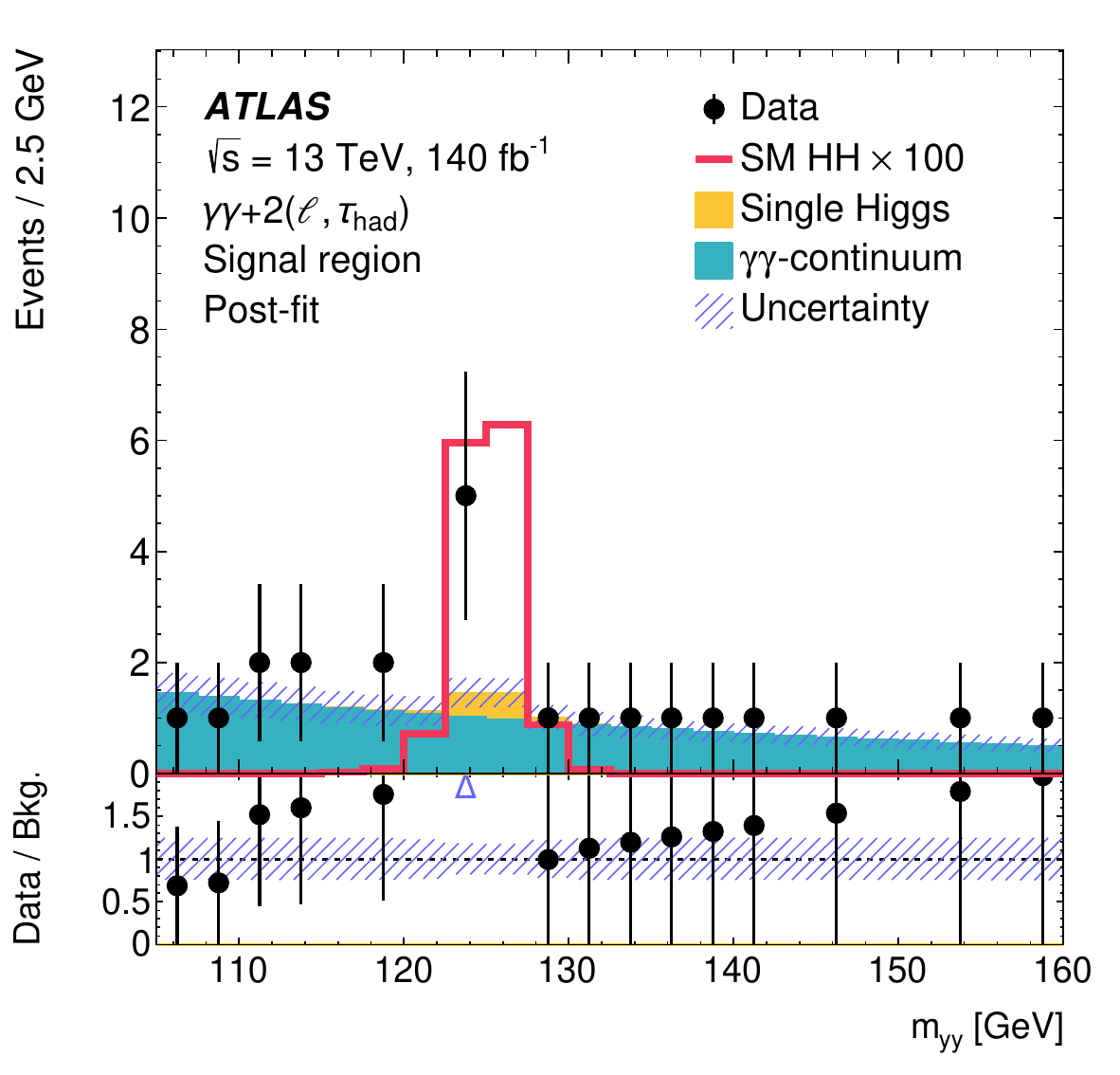}
}\\
\subfloat[]{ \includegraphics[width=0.32\textwidth]{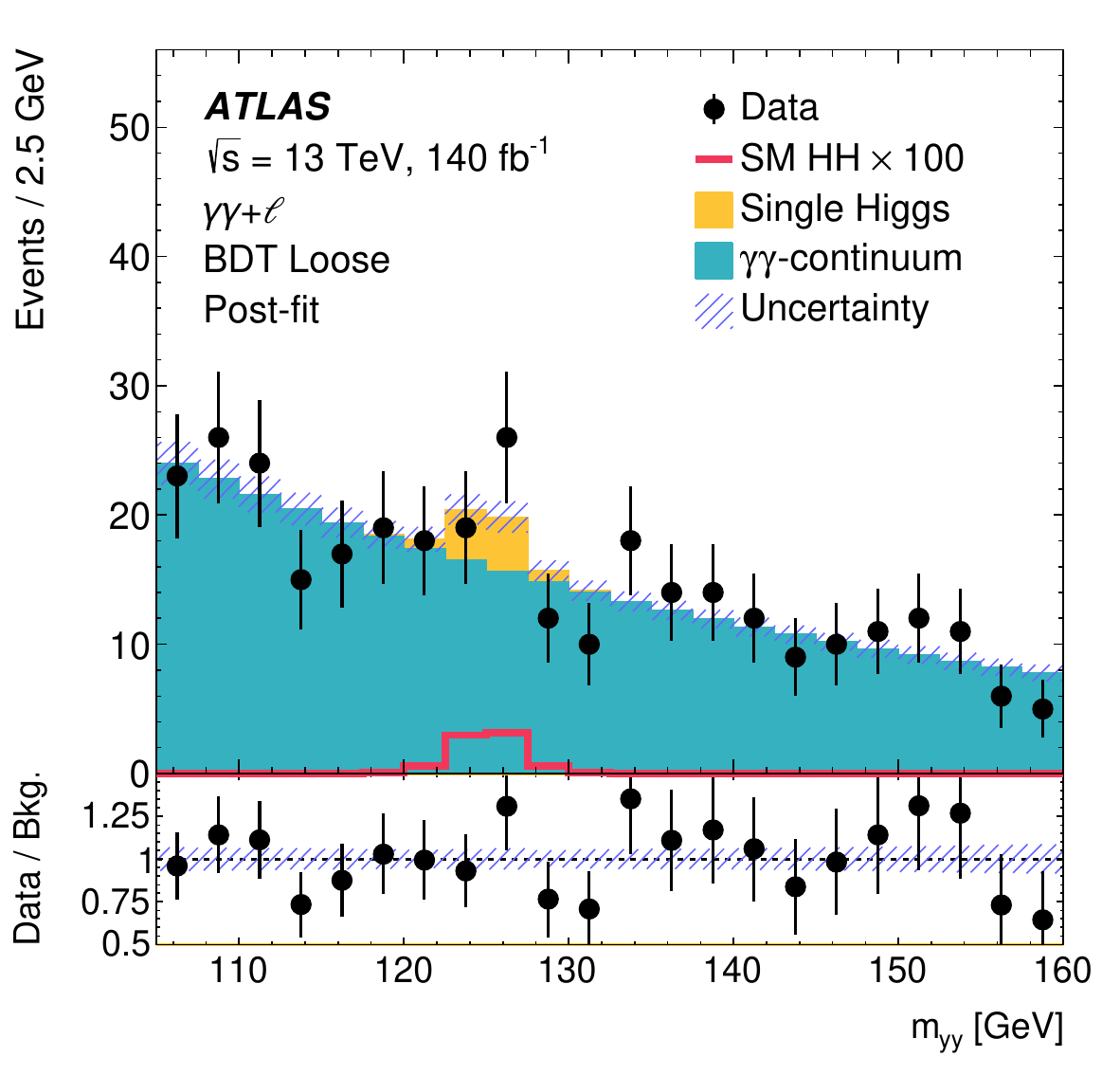}
}
\subfloat[]{ \includegraphics[width=0.32\textwidth]{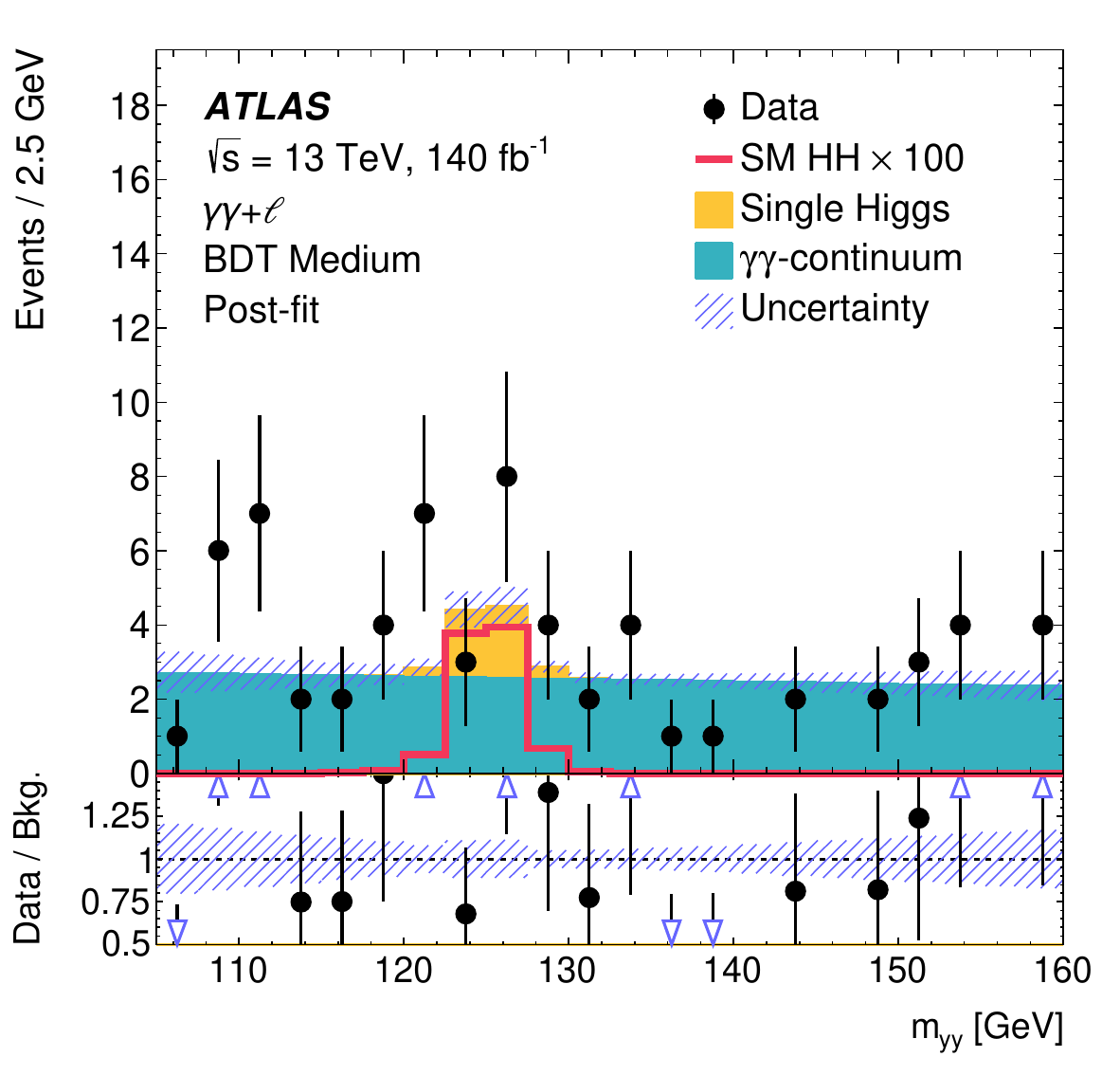}
}
\subfloat[]{ \includegraphics[width=0.32\textwidth]{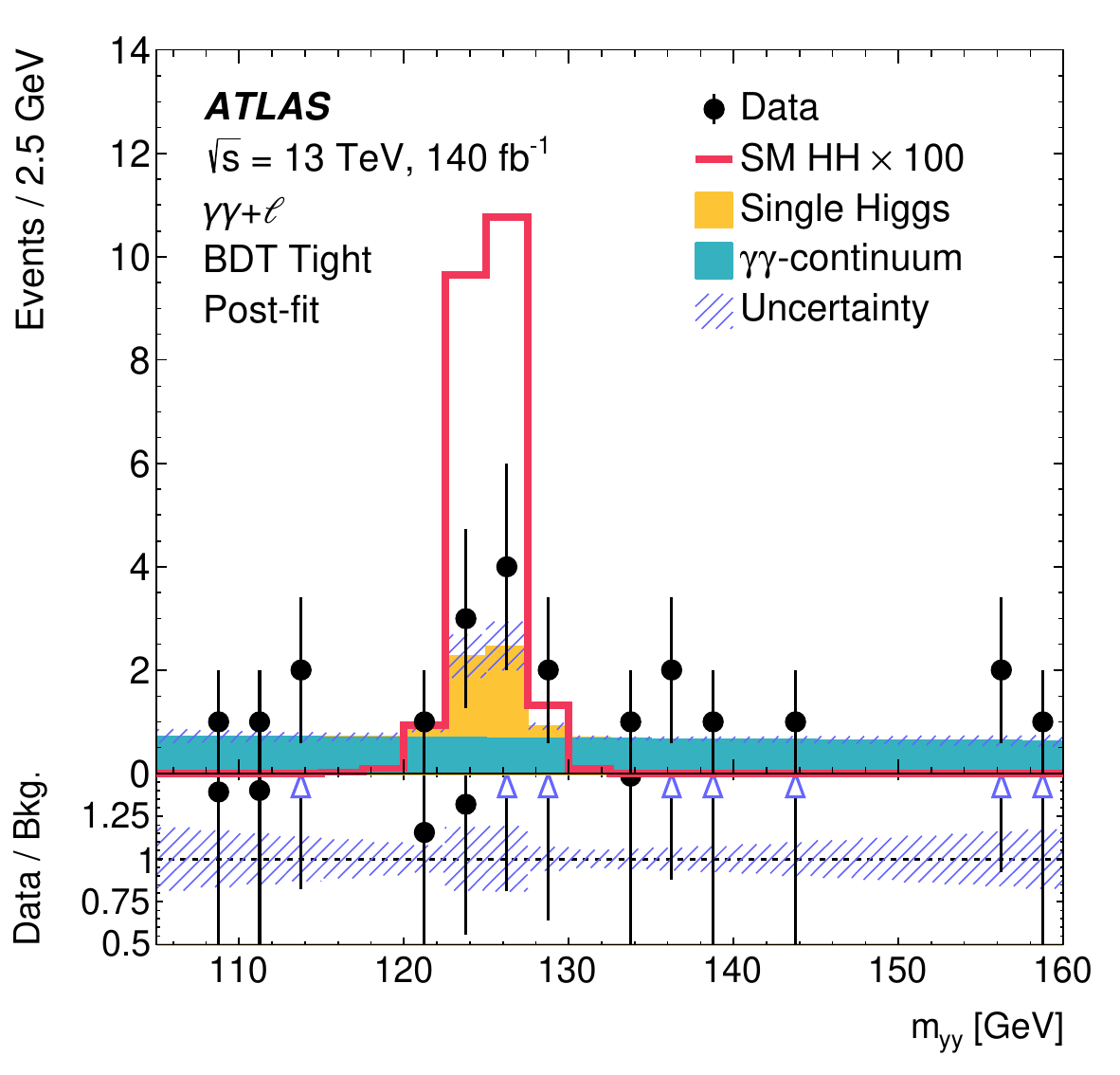}
}\\
\subfloat[]{ \includegraphics[width=0.32\textwidth]{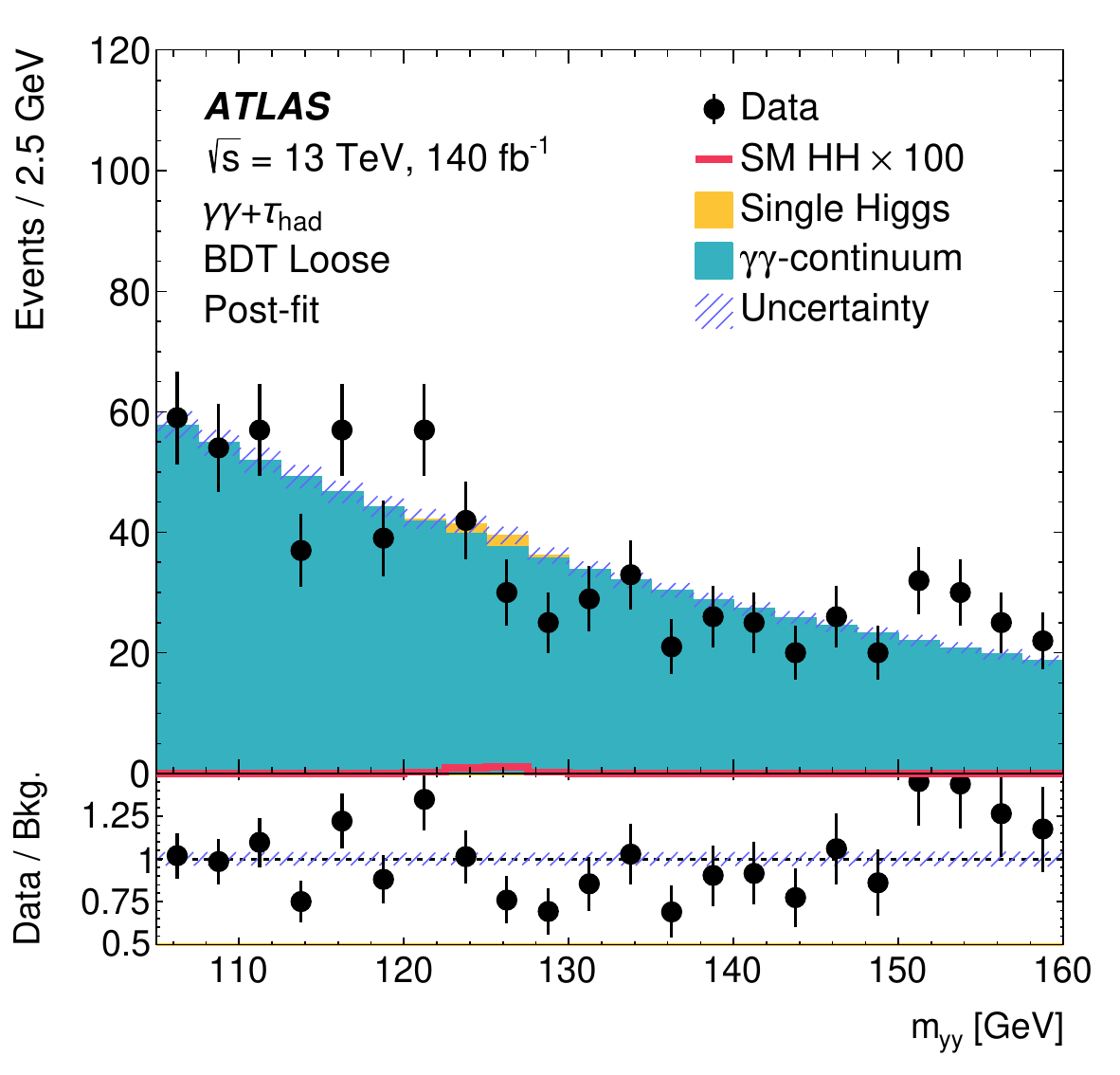}
}
\subfloat[]{ \includegraphics[width=0.32\textwidth]{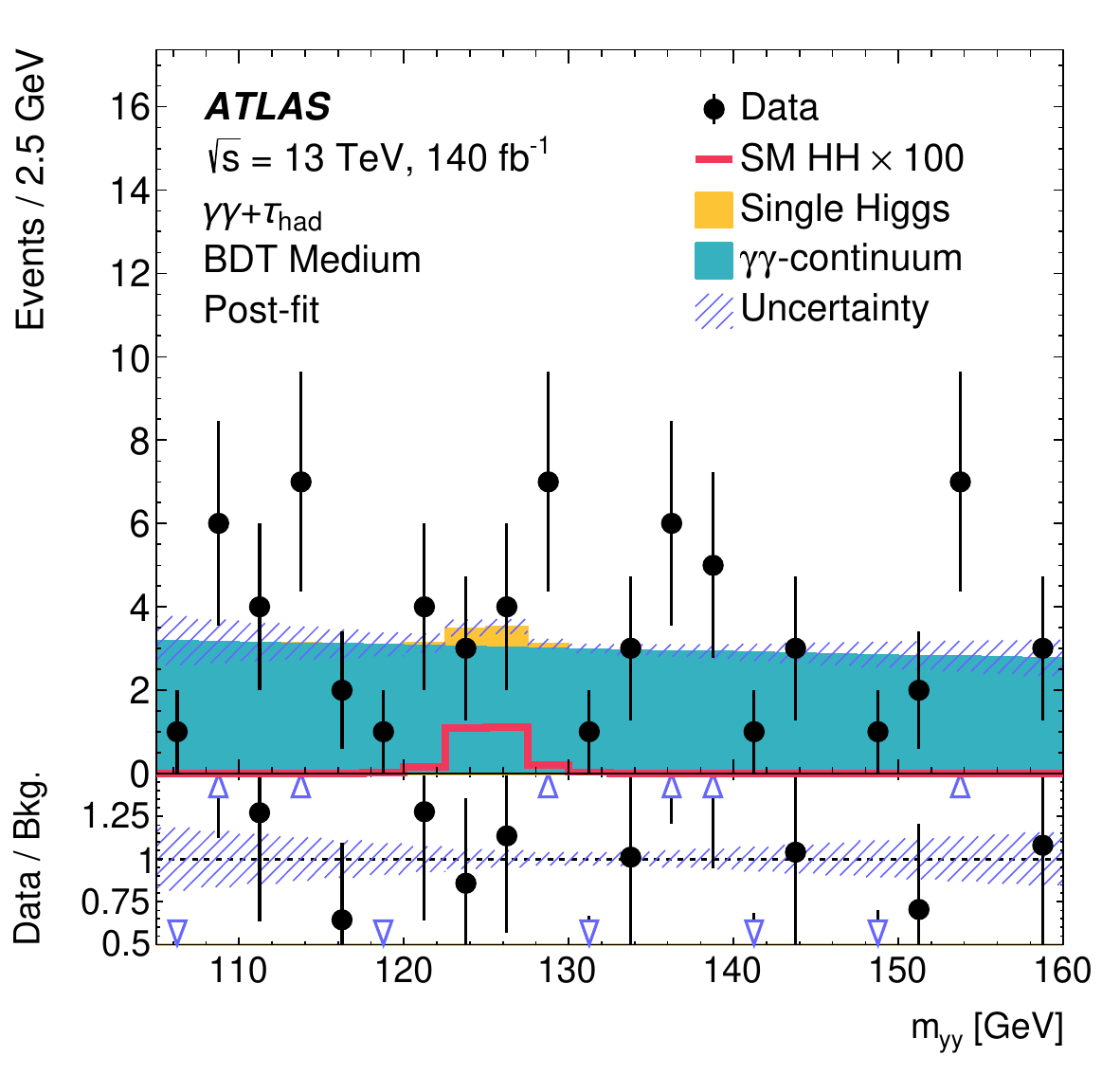}
}
\subfloat[]{ \includegraphics[width=0.32\textwidth]{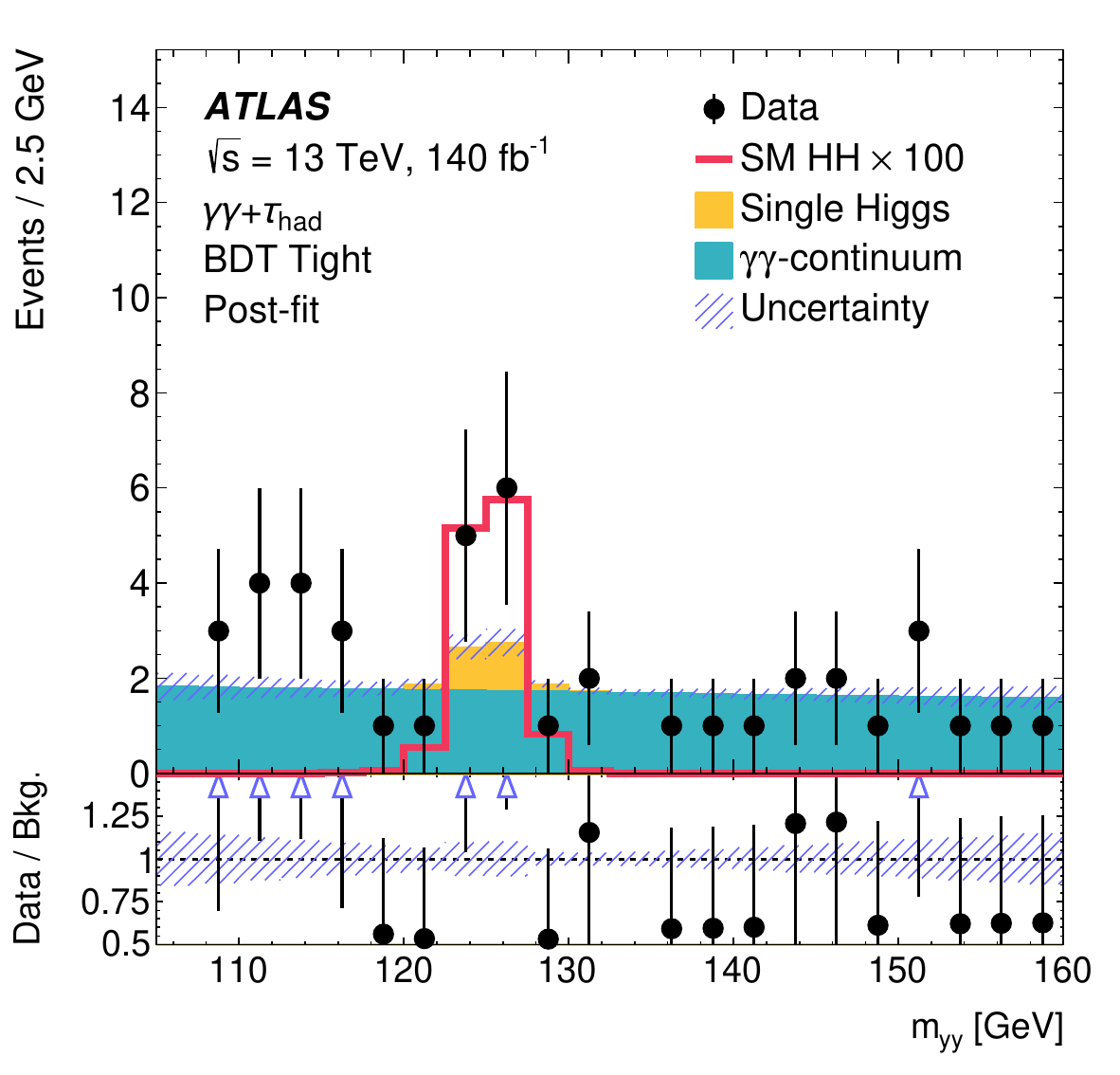}
}
\caption{Distributions of the invariant mass of the diphoton system
in the (a)~\yyll, (b--d)~\yyl, and (e--g)~\yytau\ channels, after
applying the selection requirements described in
Section~\ref{sec:events}\ and the likelihood fit to data under the
background-only hypothesis as described in
Section~\ref{sec:results}.  The \yyl\ and \yytau\ channel
distributions are shown separately for the (b,e) Loose, (c,f)
Medium, and (d,g) Tight signal regions.  The SM \HH\ signal scaled
up by a factor of 100 is also shown.  The uncertainty bands include
all sources of statistical and systematic uncertainties in the
background prediction. }
\label{fig:results_yyml}
\end{figure}


\section{Background estimation }
\label{sec:backgrounds}

 
The background composition varies for the different channels.
Processes where the event selection requirements described in
Section~\ref{sec:events}\ are satisfied by prompt leptons and real
hadronic taus produced in the decay chain are estimated using MC
simulations (described in Section~\ref{subsec:mc}).  Of these, the
dominant background processes are normalised using control regions
(CRs) in data that are orthogonal to the signal regions.
Normalisation factors are derived by performing a simultaneous fit of
all CRs and the signal region for each channel, as described in
Section~\ref{sec:results}.  Contributions from processes where at
least one of the candidate leptons is a non-prompt lepton from
$b$-hadron decays or photon conversions, or a lepton with
misidentified charge, are estimated using template fits of simulated
samples to data in CRs.  Jets misidentified as \tauhadvis\ are
referred to as fake-\tauh\ and are also estimated using data-informed
corrections to simulations, as are the non-resonant \yy\ processes
that constitute the dominant background in the \yyML\ channels.  All
ML channels use a validation region (VR) to verify the background
modelling outside of the signal region and good agreement between data
and the background predictions is observed throughout.  The
requirements that are applied to define the various control and
validation regions, relative to the preselection requirements defined
in Section~\ref{sec:events}, are shown in Table~\ref{tab:crvr_lep}\
for channels without \tauhadvis, and Table~\ref{tab:crvr_tau}\ for
channels with \tauhadvis.  Details about how the different types of
background processes are estimated are given below. The evaluation of
the systematic uncertainties associated to the (semi-)data-driven
background is also described.  Theoretical uncertainties in MC
simulations are detailed in Section~\ref{subsec:uncertainties_theory}.
 
\begin{DIFnomarkup}
\begin{table}[!tbhp]
\begin{center}
\caption{Selection criteria applied to form the control and
validation regions used to estimate backgrounds, for
the \bbllll, \lll, and \SSll\ channels relative to those used to
define the preselection regions in Table~\ref{tab:sr_defs_ml}.
Requirements that are unchanged with respect to the preselection
region are not listed (and indicated with a `--' if completely
unchanged for a given type of object). The multiplicity of jets, and
of $b$-jets are denoted \Njet\ and \Nbjet, respectively.  When no \pt\
(or \ET) threshold is specified, the default requirements for each
object are used, as described in Section~\ref{sec:objects}.
Same-charge (opposite-charge) requirements between objects are
denoted by `SC' (`OC').  The notation `SF' is used to indicate where
leptons are required to have the same flavour.  SFOC (SFSC) stands
for same-flavour, opposite-charge (same-flavour, same-charge).  The
`\Zveto' requires that the invariant mass of two SFOC leptons must
satisfy $|m_{\ell\ell}-m_{Z}| > 10$\,\GeV, while `\Zreq' inverts
this selection. In the \bbllll\ channel, the SFOC lepton pair with
an invariant mass closest to the $Z$\ boson mass is defined as the
lepton pair coming from the on-shell $Z$\ boson decay
(on-shell-$\ell\ell$) while the remaining SFOC lepton pair is
defined as coming from the off-shell $Z$\ boson decay
(off-shell-$\ell\ell$).  The radius of a conversion vertex from the
primary vertex is denoted by \rvtx, and the invariant mass of the
two opposite-charge tracks at the conversion vertex by \mtrktrk.
Regions that are included in the final fit are indicated with
a \star.
}
{\small
\begin{tabular}{l|l|cccc} \toprule
Channel      & Region                 & Leptons                            & Jets               & \bjets         & Additional                    \\
&                        &                                    &                    &                & selections                    \\ \midrule
\bbllll      & \ttbar\ CR\star        & Off-shell-$\ell\ell$\ not SFOC     & --                 & --             & --                            \\
&                        & \Zveto                             &                    &                &                               \\
& \ttZ\ CR\star          & Off-shell-$\ell\ell$\ not SFOC     & --                 & --             & --                            \\
&                        & All $\ell$\ pass loose PLV         &                    &                &                               \\
&                        & \Zreq                              &                    &                &                               \\
&                        & \mllll\ req. removed               &                    &                &                               \\
& $VV$, $H$ CR\star      & All $\ell$\ pass loose PLV         & --                 & $\Nbjet=0$     & --                            \\
& \Zjet\ CR\star         & $\pt(\ell_3) < 10$\,\GeV           &                    & --             & --                             \\
&                        & $\pt(\ell_4) < 10$\,\GeV           &                    &                &                                \\
&                        & \Zreq                              &                    &                & --                             \\
& VR                     & --                                 & --                 & --             & $|\mllll - \mH| > 10$\,\GeV    \\ \midrule
\lll         & $WZ$\ CR\star          & \Zreq                              & --                 & --             & $\met > 30$\,\GeV              \\
& HF-$e$\ CR\star        & \lscone, \lsctwo\ both $e$         & $\Njet\ge 2$       & $\Nbjet\ge 2$  &                                \\
&                        & No PLV on any $\ell$               &                    &                &                                \\
& HF-$\mu$\ CR\star      & \lscone,\lsctwo\ both $\mu$        & $\Njet\ge 2$       & $\Nbjet\ge 2$  &                                \\
&                        & No PLV on any $\ell$               &                    &                &                                \\
& Mat. conv. CR\star     & $\mlllmZ < 10$\,\GeV               & --                 & --             & --                             \\
&                        & \lscone\ or \lsctwo: \rvtx > 20\,mm&                    &                &                                \\
&                        & $0 < \mtrktrk < 100$\,\MeV         &                    &                &                                \\
& VR                     & --                                 & --                 & --             & BDT$~<~0.55$                   \\ \midrule
\SSll        & $WZ$\ CR\star          & $\ge 3\ell$(T), \pt$~>~20$\,\GeV   & --                 & --             & $\met > 30$\,\GeV              \\
&                        & One SFOC pair                      &                    &                &                                \\
&                        & \Zreq                              &                    &                &                                \\
&                        & \mll\ (any pair)~$> 12$\,\GeV      &                    &                &                                \\
&                        & $\mlllmZ~>~10$\,\GeV               &                    &                &                                \\
& $VV$jj CR\star         & \Zveto\ (SFSC pair)                & \mjj$~>~300$\,\GeV & --             & BDT$~<~-0.4$                   \\
&                        &                                    &                    &                & BDT$_{Z\text{+jets}}~>-0.8$      \\
& HF-$e$\ CR1\star       & $\ell$(T)$e$(T), no PLV            & $2 \le \Njet\le 3$ & $\Nbjet = 1$   & --                             \\
& HF-$e$\ CR2\star       & $\ell$(T)$e$(T), no PLV            & $2 \le \Njet\le 3$ & $\Nbjet\ge 2$  & --                             \\
& HF-$\mu$\ CR\star      & $\ell$(T)$\mu$(T), no PLV          & $2 \le \Njet\le 3$ & $\Nbjet\ge 1$  & --                             \\
& Mat. conv. CR\star     & \rvtx > 20\,mm                     & --                 & $\Nbjet\ge 1$  & --                             \\
&                        & $\mtrktrk < 100$\,\MeV             &                    &                &                                \\
& Int. conv. CR\star     & \rvtx < 20\,mm                     & --                 & $\Nbjet\ge 1$  & --                             \\
&                        & $\mtrktrk < 100$\,\MeV             &                    &                &                                \\
& $Q$\ mis-ID CR         & $2e$(T), OC or SC                  & $\Njet < 2$        & --             & --                             \\
& VR                     & --                                 & --                 & --             & BDT$~<~-0.4$                   \\ \bottomrule
\end{tabular}
}
\label{tab:crvr_lep}
\end{center}
\end{table}
\end{DIFnomarkup}
 
\begin{DIFnomarkup}
\begin{table}[!tbhp]
\begin{center}
\caption{Selection criteria applied to form the control and
validation regions used to estimate backgrounds, for
the \SSlltau, \lltautau, and \ltautau\ channels relative to
those used to define the preselection regions in
Table~\ref{tab:sr_defs_ml}. Requirements that are unchanged with
respect to the preselection region are not listed (and indicated
with a `--' if completely unchanged for a given type of object). The
multiplicity of (anti-ID)\tauhvis, jets, and $b$-jets are denoted
$N_{\text{(anti-ID)}\tau}$, \Njet, and \Nbjet, respectively.  When
no \pt\ (or \ET) threshold is specified, the default requirements
for each object are used, as described in Section~\ref{sec:objects}.
Same-charge (opposite-charge) requirements between objects are
denoted by `SC' (`OC').  The notation `SFOC' stands for
same-flavour, opposite-charge. The `\Zveto' requires that the
invariant mass of two SFOC leptons must satisfy
$|m_{\ell\ell}-m_{Z}| > 10$\,\GeV, while `\Zreq' inverts this
selection.  Regions that are included in the final fit are indicated
with a \star.  The \lltautau\ and \ltautau\ channels use the same
regions of data to derive fake-factors in the \Zjet-enriched and
\ttbar-enriched CRs.
}
{\small
\begin{tabular}{l|l|ccccc} \toprule
Channel        & Region                       & Leptons                     & (anti-ID) \tauhadvis                 & Jets          & \bjets         & Additional     \\
&                              &                             &                                      &               &                & selections     \\ \midrule
\SSlltau       & $VV$\ CR\star                & --                          & --                                   & --            & --             & BDT$~<~-0.2$   \\
& HF-$e$\ CR1\star             & $\ell$(T)$e$(T), no PLV     & --                                   & $\Njet\ge 2$  & $\Nbjet = 1$   & --             \\
& HF-$e$\ CR2\star             & $\ell$(T)$e$(T), no PLV     & --                                   & $\Njet\ge 2$  & $\Nbjet\ge 2$  & --             \\
& HF-$\mu$\ CR\star            & $\ell$(T)$\mu$(T), no PLV   & --                                   & --            & --             & --             \\
& Fake-\tauhadvis\ CR          & OC leptons                  & --                                   & --            & --             & --             \\
&                              & \Zveto                      &                                      &               &                &                \\
& \Zjet\ VR                    & OC leptons                  & --                                   & --            & --             & --             \\
&                              & \Zreq                       &                                      &               &                &                \\
& \ttbar\ VR                   & OC leptons                  & --                                   & $\Njet = 2$   & $\Nbjet = 1$   & --             \\
&                              & \Zveto                      &                                      &               &                &                \\
& VR                           & --                          & --                                   & $\Njet < 2$   & --             &                \\ \midrule
\lltautau      & \Zjet\ CR                    & 2$\ell$(T), OC              & $\Ntau + \Nantitau = 2$              & $\Njet \ge 1$ & $\Nbjet = 0$   & --             \\
and            &                              & \Zreq                       &                                      &               &                &                \\
\ltautau       & \ttbar\ CR                   & 2$\ell$(T), OC              & $\Ntau + \Nantitau = 2$              & $\Njet \ge 1$ & $\Nbjet = 1$   & --             \\
&                              & \Zveto                      &                                      &               &                &                \\ \midrule
\lltautau      & Fake-\tauhadvis\ CR          & --                          & ($\Ntau = 1$, $\Nantitau = 1$)       & --            & --             & --             \\
&                              &                             & or $\Nantitau = 2$                   &               &                &                \\
& Fake-\tauhadvis\ VR          & --                          & SC \tauhadvis                        & --            & --             & --             \\ \midrule
\ltautau       & Fake-\tauhadvis\ CR          & --                          & ($\Ntau = 1$, $\Nantitau = 1$)       & --            & --             & --             \\
&                              &                             & or $\Nantitau = 2$                   &               &                &                \\
& Fake-\tauhadvis\ VR          & --                          & SC \tauhadvis                        & --            & --             & --             \\ \bottomrule
\end{tabular}
}
\label{tab:crvr_tau}
\end{center}
\end{table}
\end{DIFnomarkup}
 
\subsection{Prompt leptons}
 
Diboson processes are a major background process to all of the ML
channels, particularly the \lll, \SSll, and \SSlltau\ channels where
these constitute approximately half the total background.
A CR requiring \met~$> 30$\,\GeV\ and selecting events that contain a
same-flavour, opposite charge pair of leptons with an invariant mass
consistent with the $Z$ boson mass, is used in the \lll\ channel to
provide a region enriched in $WZ$\ events.  Normalisation factors
based on the jet multiplicity are calculated by comparing simulation
to data in the CR and then applying the derived normalisation factors
to simulated events in the signal region.  Events with four or more
jets are treated inclusively.  The statistical uncertainty on the
normalisation factor in each bin is taken as the systematic
uncertainty associated to the method.  The normalisation factors, $\mu$,
range from 0.92$\pm$0.09 for events with one jet, to 0.75$\pm$0.15
for events with four or more jets.
Two CRs are employed in the \SSll\ channel to normalise diboson
processes, one enriched in $WZ$\ events ($WZ$\ CR) that follows
closely the definition used for the equivalent CR in the \lll\
channel, and the other targeting $VV$\ production in association with
two or more jets ($VV$jj CR), which controls the significant
background from vector boson scattering (VBS) processes, and is
dominated by the same-charge $W$ boson pairs component (VBS
$W^{\pm}W^{\pm}$).  The full definitions are provided in
Table~\ref{tab:crvr_lep}.  The $WZ$ CR corrects for mismodellings in
the MC simulations for diboson events with large jet
multiplicity~\cite{ATL-PHYS-PUB-2017-005}, while the $VV$jj CR
corrects for known mismodellings in the simulation of VBS
processes~\cite{STDM-2017-06}.  The \HT\ (scalar sum of the \pt\ of
all jets) distribution in the $VV$jj CR is included in the final fit,
as is a single bin in the $WZ$\ CR where only the overall
normalisation is considered and no corrections are made to the shape
of the MC simulation.
Normalisation factors of
0.79$\pm$0.05 and 1.61$\pm$0.13 are obtained for the $WZ$\ and VBS
$W^{\pm}W^{\pm}$ processes, respectively.
A low-BDT-score ($< -0.2$) region is used to constrain the $VV$\
background in the \SSlltau\ channel.  This region is included in the
final fit and a $\mu$\ of 0.91$\pm$0.23 is
obtained.
 
A CR requiring that there are no $b$-jets in the event and that all
four leptons satisfy the isolation requirements is used in the \bbllll\
channel to simultaneously constrain the $VV$\ and single Higgs
backgrounds in the fit.  Normalisation factors of 1.12$\pm$0.46 and
1.09$\pm$0.42 are obtained.
Other CRs, defined in Table~\ref{tab:crvr_lep}, are used in
the \bbllll\ channel to constrain the \ttZ\
($\mu$~=~1.27$\pm$0.22), \ttbar\ ($\mu$~=~1.50$\pm$0.28) and \Zjet\
($\mu$~=~1.01$\pm$0.36) backgrounds.  The latter two contain a mix of
prompt and non-prompt or misidentified leptons but no attempt is made
to separate these and the background shape is determined using MC
simulations. The BDT classifier used to discriminate signal from
background is applied to the data and simulated samples in
the \bbllll\ channel CRs and the resulting BDT output score
distributions are included in the final fit.
 
The normalisation of the \ttW\ background in the \SSll\ channel is
obtained while performing the final fit by allowing the normalisation
of the \ttW\ process to float in the three CRs used to constrain the
non-prompt lepton backgrounds from decays of heavy-flavour hadrons, as
described below. The \ttW\ process, which can yield a true same-charge
lepton pair, provides a significant contribution in these CRs.  A
$\mu$\ of 1.17$\pm$0.34 is obtained.  For all other
cases, background processes involving prompt leptons and real hadronic
taus are taken directly from MC simulations and normalised to their
cross-sections at the highest order available.
 
\subsection{Non-prompt leptons}
The non-prompt lepton background category encompasses events where
lepton candidates do not originate from the primary interaction point.
These non-prompt lepton backgrounds arise from various sources including
\ttbar, \Zjet, \Wjet, \Vy, and other processes
where a lepton is produced from a heavy-flavour ($b$, $c$) hadron
decay or from photon conversions.  Non-prompt leptons contribute a
significant source of background in the \lll, \SSll, and \SSlltau\
channels and are estimated by using a template fit method where a
simultaneous fit of the MC simulations to data is performed in several
CRs (and the signal regions), each enriched in a different source of
non-prompt leptons.
 
Photon conversions, where high-energy photons transform into
electron-positron pairs, occur through two primary mechanisms:
internal conversions and material conversions. Internal conversions
stem from electron-positron pair creation in a decay that might
otherwise have emitted a photon.  Material conversions occur when
high-energy photons interact with detector materials, generating
electron-positron pairs within the detector.
In the \lll\ channel, the contribution of non-prompt leptons coming
from internal conversions is very small and the shape and
normalisation of this background is estimated using MC simulation.
Conversely, the material conversion backgrounds in the \lll\ channel
are constrained using a control region (`Mat. conv. CR' in
Table~\ref{tab:crvr_lep}) where the same-charge leptons are
required to be associated with a conversion vertex with radius
$\rvtx > 20$\,mm from the primary vertex, and the invariant mass of the two
tracks at the conversion vertex, $\mtrktrk < 100$\,\MeV. An additional
requirement that the invariant mass of the three leptons be consistent
with the $Z$ boson mass is also applied to preferentially select
$Z\rightarrow \ell\ell\gamma^{*}(\gamma^{*}\rightarrow\ell^{'}\ell^{'})$\
events where one of the leptons is out of acceptance, and further
enrich the region with events containing a material conversion.  This
region is included in the final fit with one bin and a $\mu$\ of
0.66$\pm$0.13 is obtained.
For the \SSll\ channel, an internal conversion CR (`Int. conv. CR' in
Table~\ref{tab:crvr_lep}) is defined by requiring that the leptons are
associated with a conversion vertex with radius $\rvtx <20$\,mm from
the primary vertex, and an invariant mass of the two opposite-charge
tracks at the primary vertex, $\mtrktrk < 100$\,\MeV.  Another CR
requiring that leptons have a conversion vertex with radius $\rvtx
>20$\,mm from the primary vertex and an invariant mass of the two
opposite-charge tracks at the conversion vertex,
$\mtrktrk < 100$\,\MeV is used to constrain material conversions (`Mat. conv. CR'
in Table~\ref{tab:crvr_lep}).  
These regions are included in the final fit with one bin and $\mu$s of
1.84$\pm$0.28 and 1.30$\pm$0.39 are obtained for the internal
conversion and material conversion processes, respectively.  Unlike in
the \lll\ channel where material conversions stem from \Vy\ and \Zjet\
events, in the \SSll\ channel this originates mostly from top quark
and \Vy\ backgrounds, which leads to the different normalisation
factors between the two channels.
Photon conversions are a negligible background in the \SSlltau\ channel
and their shape and normalisation is taken directly from simulations.
 
Additional dedicated CRs are defined to estimate backgrounds
originating from the decay of heavy-flavour hadrons into muons
(HF-$\mu$) or electrons (HF-$e$).
For the HF-$e$\ (HF-$\mu$) CR in the \lll\ channel, the same-charge
leptons \lscone\ and \lsctwo\ are chosen to be electrons (muons) and the
PLV isolation requirements are dropped for all three leptons.
Additional selections requiring that there are at least two $b$-tagged
jets are also applied to these CRs to further enrich the heavy-flavour
backgrounds.  The distribution of the \dR\ between
\loc\ and \lscone\ in the CRs are included in the final fit and
$\mu$s of 1.50$\pm$0.50 and 1.51$\pm$0.23 are
obtained from the HF-$e$ CR and HF-$\mu$ CR, respectively.
Two CRs requiring at least one electron, removing the requirements on
the PLV isolation, and requiring that there are two or three jets in
the event, are employed in the \SSll\ channel to estimate the HF-$e$
background.  The first CR requires that exactly one of the jets is
$b$-tagged, while the second requires that exactly two jets are
$b$-tagged.  The two HF-$e$\ CRs contain different relative
contributions from electrons with a mis-ID charge (discussed in
Section~\ref{subsec:bkg_charge_mis_id}\ below) and including both CRs
provides additional information that improves the constraints on the
HF-$e$\ background.  An analogous region is used as a HF-$\mu$ CR,
where events with at least one $\mu$ are selected with the same lepton
isolation requirements as in the HF-$e$ CR.  In addition, the HF-$\mu$
CR is required to have two or three jets, at least one of them
$b$-tagged.  The distribution of the $\Delta R$ between the two
leptons in the first HF-$e$\ CR, and the distributions of the scalar
sum of the $\pt$ of all leptons and \met\ in the second HF-$e$\ CR
and the HF-$\mu$\ CR, are included in the final fit.  Normalisation
factors of 1.17$\pm$0.30 and 1.63$\pm$0.20 are obtained for the
HF-$e$\ and HF-$\mu$ backgrounds, respectively.
For the \SSlltau\ channel, the same CR definitions as for the \SSll\ channel
are applied relative to the preselection requirements, but loosening
the jet multiplicity requirements to allow events to have at least two jets.
Normalisation factors of 0.87$\pm$0.09 and 0.75$\pm$0.06
are obtained for the HF-$e$\ and HF-$\mu$\ backgrounds, respectively.
 
The systematic uncertainty in this template fit method for the various
non-prompt lepton background components is determined by relaxing the
isolation and identification criteria applied to the leptons.  The
templates obtained using MC simulations are compared with the shape of
distributions in data after subtracting the expected contributions
from processes with prompt leptons using MC simulations.  For each
source of non-prompt lepton backgrounds, the difference between the
simulation-based template and these residual data events obtained with
the adjusted criteria are considered as uncertainties in the shape of
the estimates obtained under the nominal conditions.
 
\subsection{Charge misassignment}
\label{subsec:bkg_charge_mis_id}
Backgrounds where the charge of the lepton was incorrectly
assigned primarily affect the \SSll\ channel.  Such events originate
from \Zjet, \ttbar\ and $WW$\ processes, where one electron undergoes
a hard bremsstrahlung and asymmetric conversion ($e^{\pm} \rightarrow
e^{\pm}\gamma^{*} \rightarrow e^{\pm}e^{+}e^{-}$), or the track
curvature is mismeasured.  The rate of electron charge mismeasurement
is measured in data by taking the ratio of same-charge and
opposite-charge pairs of electrons in a high purity sample of
\Zepem\ events, following the method described in
Ref.~\cite{EGAM-2018-01}. The CR is defined by selecting events
satisfying the \SSll\ preselection criteria, but removing the lepton
charge requirements and requiring that there are less than two jets in
the event.  The charge misidentification rates are measured separately
for prompt electrons and electrons that originate from either an
internal conversion or a material conversion (following analogous
selection requirements as used for the non-prompt lepton CRs described
above).  Rates are calculated as a function of \pt\ and $\eta$\ of the
electrons in each category and range from $10^{-5}$\ for low-\pt\
prompt electrons to $10^{-1}$ for high-\pt\ electrons with a
large-radius conversion, but are more typically around $10^{-3}$.  The
measured charge misassignment rate is applied to data events
satisfying the requirements of the \SSll\ channel preselection, but
requiring that the two leptons have opposite charge.  Uncertainties in
the method are evaluated by comparing the nominal rates with those
computed using simulated \Zepem\ events, and by varying the requirements
on the dielectron invariant mass used to select \Zepem\ events.  The
muon charge misassignment rate is negligible in the \pt\ range
considered. Processes with a misassigned charge
constitute less than 1\% of events in the \lll\ and \SSlltau\ signal
regions and for these channels their contribution is taken from the MC
simulations.
 
\subsection{Misidentified hadronic taus}
\label{subsec:background_taus}
 
Quark- or gluon-initiated jets that are incorrectly reconstructed as
a \tauhadvis\ (fake-\tauhadvis) are an important background in
the \SSlltau, \lltautau, and \ltautau\ channels.  In the \SSlltau\
channel, background processes where a jet fakes the \tauhadvis\ are
estimated by deriving scale factors to correct the rate of jets to be
misidentified as hadronic taus in MC to match the rate in data.  The scale
factors are derived by comparing the rates of jets satisfying
the \tauhadvis\ identification requirements in data, to the rate in
MC simulations, in a control region defined by applying the same preselection
requirements described in Section~\ref{sec:events}\ but requiring that
the two light leptons have opposite-sign charge and that their invariant
mass is not compatible with \mZ.  Contributions from processes
containing real \tauhadvis\ or prompt leptons are subtracted from data
using predictions from MC simulations before computing the ratio.  Scale factors are
derived separately for one- and three-prong taus, as a function of
the \tauhadvis \pt.  The derived scale factors are applied to the
relevant simulated events in the signal region and are in the range of
0.68--0.86 (0.48--0.82) for one-prong (three-prong) taus.  Two VRs,
enriched with fake-\tauhadvis\ in \Zjet\ and \ttbar\ events are
respectively defined by modifying the CR to require that the
invariant mass of the two light leptons is consistent with the $Z$\ boson mass, and by requiring
that there are exactly two jets in the event, exactly one of which
passes the $b$-tagging requirements.  The largest difference in each
region and \pt\ bin, between the scale factors derived in the nominal
CR and those derived in the VRs, is taken as an uncertainty in the
method.  An additional source of systematic uncertainty is considered
by varying the real-\tauhadvis\ contribution from simulations up and down by
50\%.  The total uncertainty in the scale factors ranges from 20\% to
34\%, depending on the \pt\ range and number of prongs considered.
 
Fake-\tauhadvis\ backgrounds in the \lltautau\ and \ltautau\ channels
are estimated from data using the fake-factor method described in
Ref.~\cite{HDBS-2022-44}. The fake-factors are estimated in a CR
enriched in \Zjet\ events (the `\Zjet\ CR' in Table~\ref{tab:crvr_tau})
that is common to both channels.  The CR requires
that there are exactly two tau candidates, each of which is required
to satisfy either the \tauhadvis\ or anti-ID~\tauhadvis\ criteria.
The events are divided into sub-regions based on whether the leading tau candidate
satisfies the \tauhadvis\ or anti-ID~\tauhadvis\ requirements, and the
fake-factors are taken as the ratio of the number of events in each
sub-region, and are derived as a function of \pt, $|\eta|$\ and number of
prongs of the (anti-ID)~\tauhadvis.  The process is repeated,
subdividing events based on whether the subleading tau satisfies
the \tauhadvis\ or anti-ID~\tauhadvis\ requirements.  Finally, the
fake-tau backgrounds are estimated in the \lltautau\ and \ltautau\
signal regions by using the derived fake-factors to reweight templates
obtained from data in CRs (the `Fake-\tauhadvis\ CR's in
Table~\ref{tab:crvr_tau}), by applying the respective signal region
requirements but requiring that at least one of the two taus instead
satisfies the anti-ID~\tauhadvis requirements.
The fake-factors are also estimated in a CR enriched in \ttbar\ events
in order to check the dependency of the fake-factors to light-flavour
quark, heavy-flavour quark, or gluon-initiated jets.  The measured
fake-factors in the \ttbar\ CR are consistent within statistical
uncertainties with the nominal ones, but the difference ($\sim$30\%)
is treated as a systematic uncertainty arising from the different jet
compositions in each region.  The contribution of real \tauhadvis\
satisfying the anti-ID \tauhadvis\ requirements is varied up and down
by 15\% to account for theoretical uncertainties in these
processes, and the impact on the derived fake-factors is considered as an
additional source of uncertainty.  The fake-\tauhadvis\ background
estimate is validated in VRs that follow the signal region definition
but require that the two \tauhadvis\ have the same-sign charge.  Good
agreement between the data and the background prediction is observed
in the \lltautau\ and \ltautau\ channel fake-\tauhadvis\ VRs, within
the available statistical precision.  A 10\% discrepancy is observed
in the \ltautau\ fake-\tauhadvis\ VR, which is then conservatively
considered as an additional uncertainty in the fake-\tauhadvis\
background estimate in both the \lltautau\ and \ltautau\ channels.
 
\subsection{Non-resonant \yy\ production}
\label{subsec:background_yy}
 
Non-resonant \yy\ production originates from \yyjet, \Vyy, and \ttyy\
processes, as well as from processes where a jet is
incorrectly identified as a photon.  This \yy-continuum background is
expected to have a smoothly falling shape.  It is modelled using a
functional form chosen by fitting the diphoton invariant mass
distribution in sidebands around the Higgs boson mass [$105\,\GeV
< \myy < 120\,\GeV$, $130\,\GeV < \myy < 160\,\GeV$] in a CR in data,
following the methodology described in
Refs.~\cite{HIGG-2013-08,HDBS-2021-10}.  The CR is defined by
requiring that events have no P-type leptons (as defined in
Table~\ref{tab:lepton_defs}) or \tauhadvis, have one (\yyl\ and \yytau\
channels) or two (\yyll\ channel) jets, and satisfy all other
preselection requirements defined in Section~\ref{sec:events}.  A
first-order exponential function is observed to provide the best fit
to the background model in all regions.  This function is used to
generate a background histogram, with floating functional form
parameters when fitting to the data in sidebands.  The fit is
performed separately in each region and each channel, but in the
\yyl\ and \yytau\ channels the BDT Medium and BDT Tight regions are
combined to obtain the fit parameters, and the obtained background
template is then normalised to the sidebands in each region separately.
 
The potential bias associated with the choice of functional form to
model the continuum background is evaluated in each
signal region by fitting the background template using a
model with free parameters following the prescriptions described in
Refs.~\cite{HIGG-2013-08, ATL-PHYS-PUB-2020-028}.  Uncertainties of up
to 4\% due to this `spurious signal' uncertainty are obtained. An
additional source of uncertainty in the shape of the non-resonant
background caused by differences in the background composition between
the signal region and the CR is estimated using MC simulations
of photon pairs produced in association with one or two jets.  The
background template is derived using these simulated samples in the CR
region, requiring exactly zero P-type leptons and \tauhadvis,
and comparing this to the background template obtained from the
simulated samples when the same lepton and \tauhadvis\ requirements of
the respective signal regions are applied.  The difference between
these two estimates is taken as the uncertainty in the nominal
background estimate derived from data.  Uncertainties in the
background normalisation of 13.1\% (8.4\%) are measured in
the Medium (Tight) BDT-score regions of the \yyl\ channel, 12.4\%
(8.0\%) in the Medium (Tight) BDT-score regions of the \yytau\
channel, and less than 2\% in all other regions.


\section{Systematic uncertainties}
\label{sec:uncertainties}

 
For every channel, the total uncertainty is dominated by the
statistical uncertainty in the number of data events in the signal
region. Experimental sources of systematic uncertainty due to the
detector response and background modelling are considered, as are
theoretical uncertainties in the normalisation and shape of signal and
background processes.  The finite statistics of MC
simulations used in the analysis are also considered as a source of
systematic uncertainty.  The impact of the different sources of
uncertainty in the expected \muHH\ upper limit at 95\% CL is
summarised in Table~\ref{tab:systs}\ for the combination of all
channnels, and for the combinations of the ML channels
and \yyML\ channels separately.

\begin{DIFnomarkup}
\begin{table}[!htb]
\begin{center}
\caption{Breakdown of the relative contributions to the
uncertainties in the expected \muHH\ upper limit at 95\% CL, as
determined in the likelihood fit to data described in
Section~\ref{sec:results}, for combinations of the ML channels,
the \yyML\ channels, and all channels.  The impact of the
uncertainties is quantified as the relative variation of the
expected upper limit when re-evaluating the profile likelihood
ratio after fixing a nuisance parameter to its best-fit value,
while all other nuisance parameters are allowed to float.
Individual sources of uncertainty that have an impact smaller
than 1\% in all channels are not listed.
}
{\small
\begin{tabular}{l c c c c c c} \toprule
\multirow{2}{*}{\shortstack[l]{Systematic\\uncertainty source}}
&\multicolumn{3}{c}{Relative impact of systematic uncertainties [\%]} \\ \cmidrule{2-4}
& ML channels          & \yyML\ channels      & Combination           \\ \toprule
\textbf{Total}                                     & 22                   & 14                   & 19                    \\ \midrule
\hspace{5mm}\textbf{MC statistics}                 & 5                    & <1                   & 3                     \\ \midrule
\hspace{5mm}\textbf{Experimental}                  & 5                    & <1                   & 3                     \\
\hspace{10mm}Detector response                     & 4                    &                      & 3                     \\
\hspace{15mm}\textcolor{gray}{ Jets and \met}      & \textcolor{gray}{3}  &                      & \textcolor{gray}{2}   \\
\hspace{15mm}\textcolor{gray}{ Flavour tagging}    & \textcolor{gray}{1}  &                      & \textcolor{gray}{<1}  \\
\hspace{10mm}Background estimate                   & <1                   & <1                   & <1                    \\ \midrule
\hspace{5mm}\textbf{Theoretical}                   & 13                   & 14                   & 13                    \\
\hspace{10mm}Signal                                & 10                   & 12                   & 11                    \\
\hspace{10mm}Backgrounds                           & 4                    & 2                    & 3                     \\
\hspace{15mm}\textcolor{gray}{ Top quark}          & \textcolor{gray}{1}  & \textcolor{gray}{--} & \textcolor{gray}{<1}  \\
\hspace{15mm}\textcolor{gray}{ Vector boson}       & \textcolor{gray}{3}  & \textcolor{gray}{--} & \textcolor{gray}{2}   \\
\hspace{15mm}\textcolor{gray}{ Single Higgs boson} & \textcolor{gray}{1}  & \textcolor{gray}{2}  & \textcolor{gray}{1}   \\
\hspace{15mm}\textcolor{gray}{ Other}              & \textcolor{gray}{<1} & \textcolor{gray}{--} & \textcolor{gray}{<1}  \\ \bottomrule
\end{tabular}
}
\label{tab:systs}
\end{center}
\end{table}
\end{DIFnomarkup}
 
\subsection{Experimental uncertainties}
 
The uncertainty in the combined 2015--2018 integrated luminosity is
0.83\%~\cite{DAPR-2021-01}, obtained using the LUCID-2 detector~\cite{LUCID2}
for the primary luminosity measurements, complemented by
measurements using the inner detector and calorimeters.  An
uncertainty arising from the correction of the pile-up distribution in
simulation to that in data is also considered.
 
The impact of uncertainties in the trigger, reconstruction, identification and
isolation efficiencies of electrons~\cite{TRIG-2018-05,EGAM-2018-01},
muons~\cite{TRIG-2018-01,MUON-2018-03}, and
photons~\cite{TRIG-2018-05,EGAM-2018-01}\ are considered.  An
additional uncertainty in the track-to-vertexing matching is applied to
muons.  Reconstruction and identification efficiency uncertainties
on \tauhadvis~\cite{ATLAS-CONF-2017-029}\ are also considered, along
with the uncertainty associated with measurements of the \tauhadvis\ energy
scale, and the efficiency of the electron veto used in the
\tauhadvis\ selection.
 
Jet energy scale and resolution uncertainties~\cite{JETM-2018-05}\ and
the uncertainty in the efficiency of matching jets to the primary
vertex~\cite{PERF-2014-03}\ are considered. These energy scale and
resolution uncertainties, in addition to an uncertainty in the tracks
matched to the primary vertex but not associated with other
reconstructed objects in the event, are propagated to the
\met\ calculation~\cite{JETM-2020-03}. Uncertainties
in the $b$-jet tagging efficiency and misidentification rates are
estimated using \ttbar\ events~\cite{FTAG-2018-01,FTAG-2020-08}\ for
$b$- and $c$-jets, and $Z$+jets events for light-flavour
jets~\cite{FTAG-2019-02}, and considered in the analysis.
 
Systematic uncertainties associated with the experimental methods used
for the background estimates are described in
Section~\ref{sec:backgrounds}.
 
\subsection{Theoretical uncertainties}
\label{subsec:uncertainties_theory}
 
Several sources of theoretical uncertainty impacting the signal models
are considered. The uncertainties linked to the modelling of the
parton shower and underlying event are assessed by comparing the
nominal sample, where the showering process is modelled
using \PYTHIA[8], with an alternative sample that uses \HERWIG[7].
Uncertainties in the matrix element calculation are assessed by
varying the factorisation and renormalisation scales employed in the
generator, either independently or concurrently, by a factor of
two. Theoretical uncertainties related to the \ggF\ \HH\
cross-section, stemming from uncertainties in the PDF and
$\alpha_{s}$\ ($\pm 3.0\%$), as well as the selection of
renormalisation scheme and the top quark mass scheme
($^{+6\%}_{-23\%}$)~\cite{Grazzini:2018bsd, Baglio:2020wgt}\ are also
considered. Uncertainties in the VBF \HH\ cross-section are also
considered and are dominated by the uncertainty in the PDF and
$\alpha_{s}$\ ($\pm 2.1\%$).  These cross-section uncertainties are
factored into the determination of the upper limits on the \HH\ signal
strength, as well as the likelihood-based constraints on the values of
the \kl\ modifier.  Theoretical uncertainties associated with the
branching ratios of the Higgs bosons~\cite{deFlorian:2016spz}\ range
from 1.2\% to 2.1\% and are also considered but their impact is negligible.  The
variation of the branching ratio uncertainty with \kl\ is not
considered.
 
Background modelling uncertainties due to the choice of generator for
the hard scatter and parton shower are considered by comparing them
with alternative simulation setups, as detailed in
Table~\ref{tab:mcconfig}, where available.  Uncertainties due to the
choice of renormalisation and factorisation scales are evaluated by
varying these by factors of 0.5 and 2, relative to the nominal scales.
For background processes where the normalisation is determined from
control regions in data, no uncertainty in the cross-section is
considered.  Uncertainties of 20\%~\cite{Frederix:2017wme}, and
5\%~\cite{TOPQ-2018-01}\ are considered on the normalisations of the
$\ttbar\ttbar$, and $t\Zy$\ processes, respectively.  Conservatively,
a 50\% uncertainty in the normalisation of $\ttbar t$, $tW$, $t\Zy$,
$t\bar t W^+ W^-$, and $VVV$\ backgrounds is applied.
 
Theoretical uncertainties associated with single Higgs boson
production cross-sections due to missing higher-order QCD corrections,
the effects of PDF and $\alpha_s$\ uncertainties, and the
uncertainties in the branching fractions, are taken from
Ref.~\cite{deFlorian:2016spz}.  The total theoretical uncertainties in
the different single Higgs boson production cross-sections are 9\%
for \ggF, 3\% for VBF, 3\% for $WH$, 4\% for $ZH$, and 11\% for \ttH.
The uncertainty in the single Higgs background processes due to the
choice of parton shower model is evaluated for the \ttH\ process in
the ML channels, and for the \ggF, VBF, $VH$\ and \ttH\ processes in
the \yyML\ channels by comparing the predictions of the nominal
simulation using the \Pythia[8]\ model with an alternative simulation
in which the same generator-level events are showered with \Herwig[7].
The uncertainties are 8\% and 10\% for the \yyML\ and \bbllll\
channels and less than 3\% for all other channels.  An uncertainty of
10\% (40\%) is assigned to cover parton shower model uncertainties on
$VH$\ backgrounds to the \lll\ and \SSll\
(\bbllll, \SSlltau, \lltautau, and \ltautau) channels, following the
observations in Refs.~\cite{HDBS-2019-27,HIGG-2018-20}.  An additional
100\% uncertainty is assigned to the \ggF, VBF, and $VH$\
processes in the \bbllll\ channel in order to account for difficulties
in the modelling of these processes in association with heavy-flavour
jets.


\section{Statistical treatment and results}
\label{sec:results}

Measurements of the \HH\ signal strength and constraints on the
self-coupling strength are obtained using a binned likelihood function
$L(\alpha, \theta)$, following the method described in
Ref.~\cite{HIGG-2012-17}.  The variable $\alpha$ represents the
parameters of interest (POI) associated with the measurement, while
$\theta$ represents nuisance parameters corresponding to the
systematic uncertainties described in Section~\ref{sec:uncertainties}
and background parameters that are constrained by control regions in
data.  Theoretical uncertainties in simulated signal and background
processes are treated as correlated across all channels, as well as
are experimental uncertainties related to the data-taking conditions
and physics objects.  Uncertainties related to background estimates
using data-informed methodologies (derived from template fits, or
estimated in CR or side-band data regions) are treated as
uncorrelated, except in cases where a common CR is used in which case
it is treated as correlated.  The global likelihood function
$L(\alpha, \theta)$ is the result of multiplying the likelihood
functions in each of the nine signal regions and the 19 CRs indicated
in Tables~\ref{tab:crvr_lep}\ and \ref{tab:crvr_tau}.  For each
channel, the likelihood function is derived from the respective signal
and background models of the probability density functions for the
variable of interest. These models take into account the expected
signal and background yields for given values of $\alpha$ and
$\theta$, and the observed distribution of the discriminating variable
in each channel -- the BDT output score distribution for each of the
ML channels, and \myy\ for the \yyML\ channels.
 
Upper limits are set on the \HH\ signal strength, \muHH\ at 95\% CL,
using the profile-likelihood-ratio test statistic and the modified
frequentist $\text{CL}_\text{s}$ technique~\cite{Read:2002hq} in the
asymptotic approximation~\cite{Cowan:2010js}.  The scenario $\muHH =
0$\ corresponds to the background-only hypothesis and $\muHH > 0$\
corresponds to the presence of an \HH\ signal in addition to the
background.  Asimov datasets~\cite{Cowan:2010js}\ are used to derive
the expected limits, with all pre-fit estimates of the nuisance
parameters set to values derived from the fit to the data, and the
parameters of interest set corresponding to the hypothesis being
tested.  The 95\% CL limits on the signal strength for individual
channels, the statistical combinations of the ML and \yyML\ signal
categories, and the combination of all channels, are shown in
Figure~\ref{fig:limits_signal-strength}.  The overall combination
yields an observed 95\% CL upper limit on \muHH\ of
\muHHobs, with an expected upper limit of
\muHHexpzero\ in the absence of \HH\ production, and
\muHHexpSM\ for the SM case.  If systematic uncertainties are neglected
then the expected limit is \muHHexpstatonly\ when assuming no \HH\
production.  The asymptotic results are found to agree within 8\% with
values obtained using pseudo-experiments.
 
\begin{DIFnomarkup}
\begin{figure}[!htb]
\centering
\includegraphics[width=0.8\textwidth]{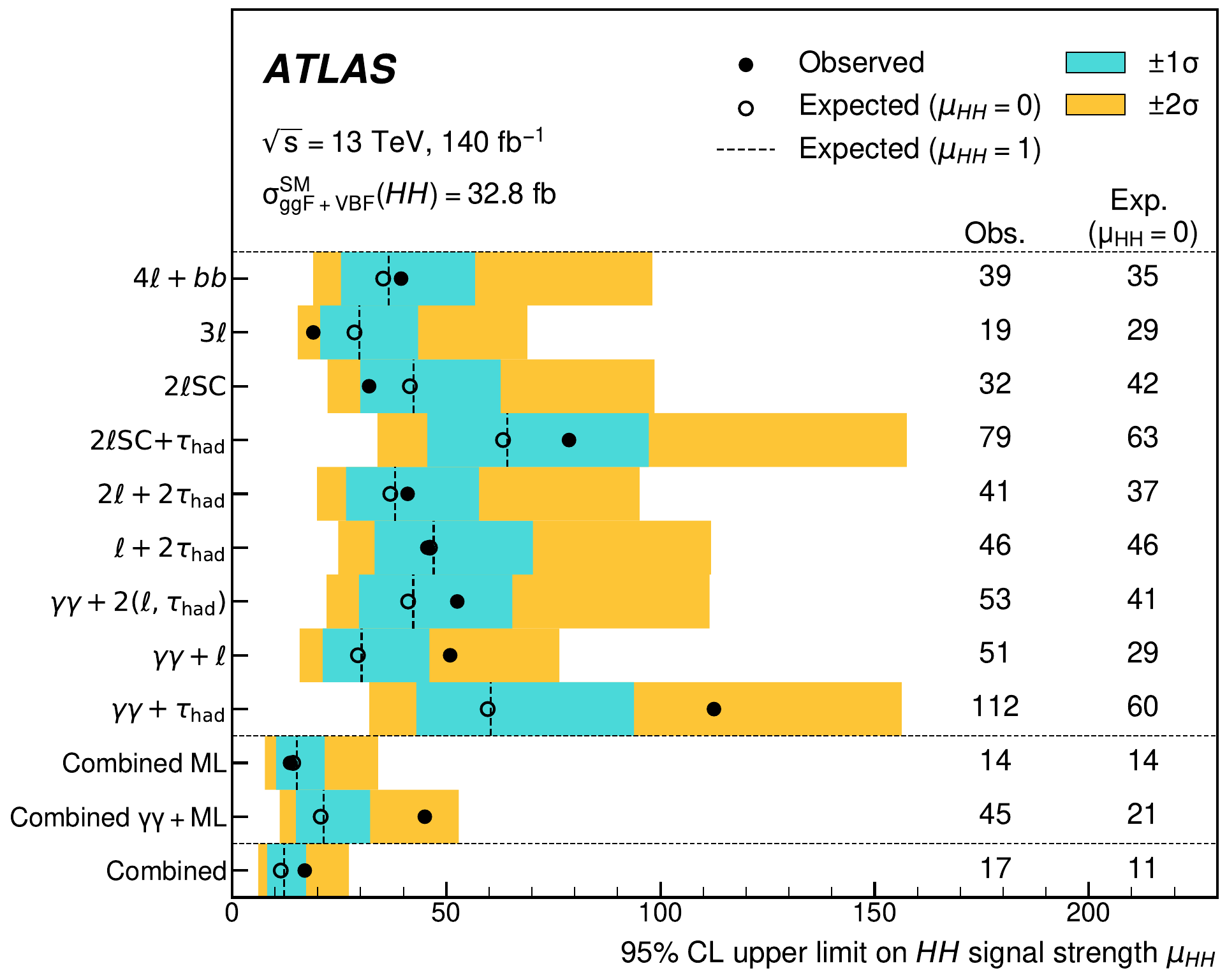}
\caption{Observed
(filled circles) and expected (open circles) 95\% CL upper limits
on the signal strength for \HH\ production in the background-only
(\muHH~=~0) hypothesis. The dashed lines indicate the expected 95\%
CL upper limits on \muHH\ in the SM hypothesis (\muHH~=~1). The
inner and outer bands indicate the $\pm 1\sigma$\ and $\pm
2\sigma$\ variations on the expected limit under the
background-only hypothesis due to statistical and systematic
uncertainties, respectively.  Results are shown individually for the
different search channels, the statistical combination of ML
and \yyML\ channels separately, and the statistical combination of
all channels.}
\label{fig:limits_signal-strength}
\end{figure}
\end{DIFnomarkup}
 
Constraints on the Higgs self-coupling strength and the $VVHH$\
coupling strength, expressed as 68\% and 95\% confidence intervals
(CIs), are determined using the method described in
Ref.~\cite{HDBS-2022-03}, using a profile-likelihood-ratio test
statistic $\Lambda(\alpha,\theta)$ computed from the likelihood
function in the asymptotic approximation~\cite{Cowan:2010js}, where
the POIs in $\alpha$ are the coupling strength modifier \kl\ and \kVV,
respectively.  The values of twice the negative-logarithm of the
profile likelihood ratio ($-2 \text{ln} \Lambda$) as a function
of \kl\ and \kVV\ are shown in Figure~\ref{fig:likelihood-scans}.  The
best-fit value of \kl\ is found to be \klbestfit\ from the profile
likelihood scan.  With the values of all other couplings fixed to
their SM value, the observed (expected) 95\% CI for \kl\ is found to
be [\klobsdown, \klobsup] ([\klexpdown, \klexpup]).  The observed
(expected) 95\% CI for \kVV\ is found to be [\kvvobsdown, \kvvobsup]
([\kvvexpdown, \kvvexpup]) with all other couplings fixed to their SM
value.  The expected limits are computed assuming the SM.  The
double-minima structure in the observed limit occurs due to a
degeneracy in the best-fit signal yield that arises because of the
competing factors of the signal cross-section and effects on the
acceptance times efficiency when varying the couplings.  As can be
seen in Figure~\ref{fig:likelihood-scans-split}, the effect is driven
by the \yyML\ channels, where a mild excess is observed in data
compared to the background prediction.
 
\begin{figure}[!tbhp]
\centering
\subfloat[]{\includegraphics[width=0.45\textwidth]{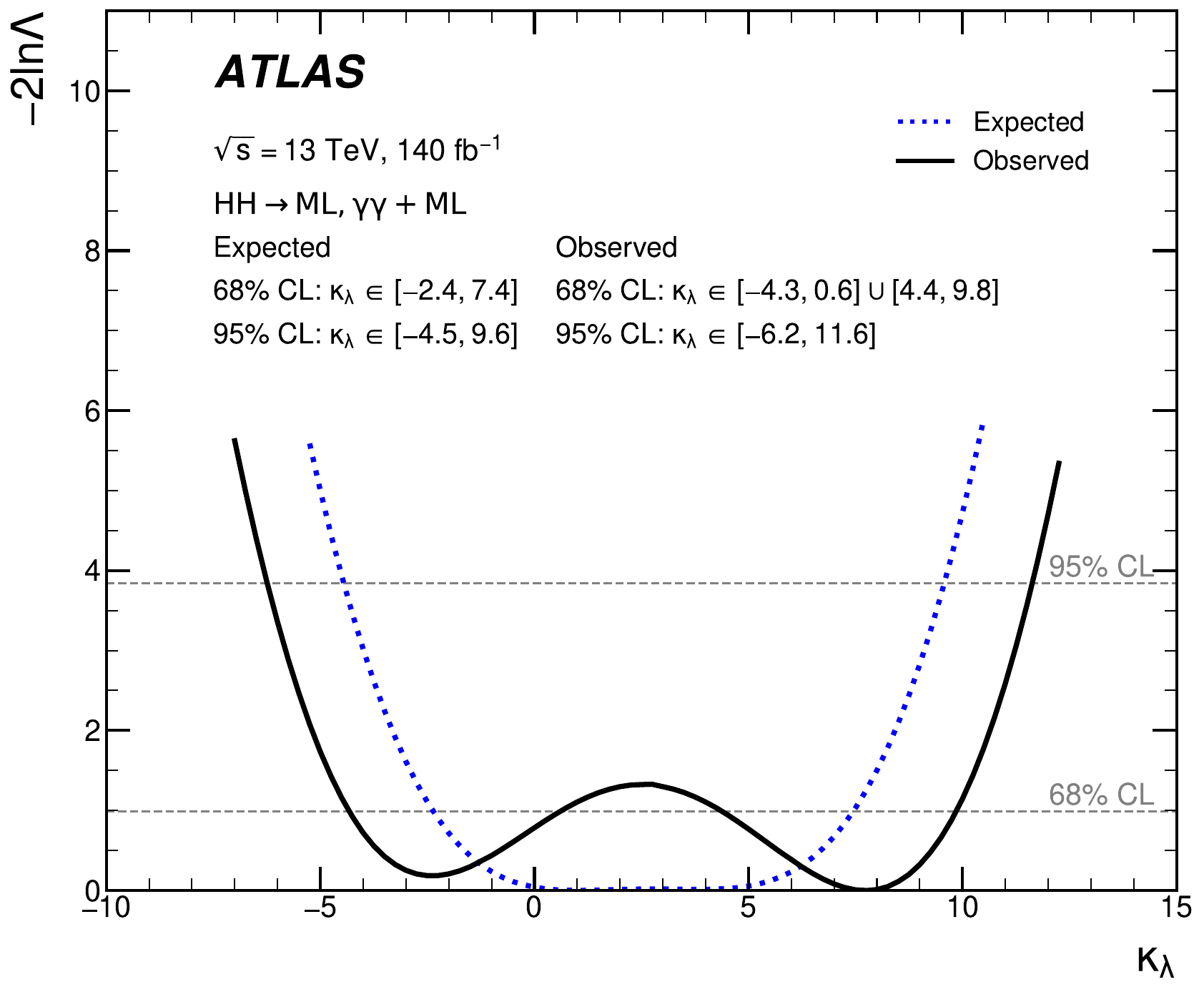}
}
\subfloat[]{\includegraphics[width=0.45\textwidth]{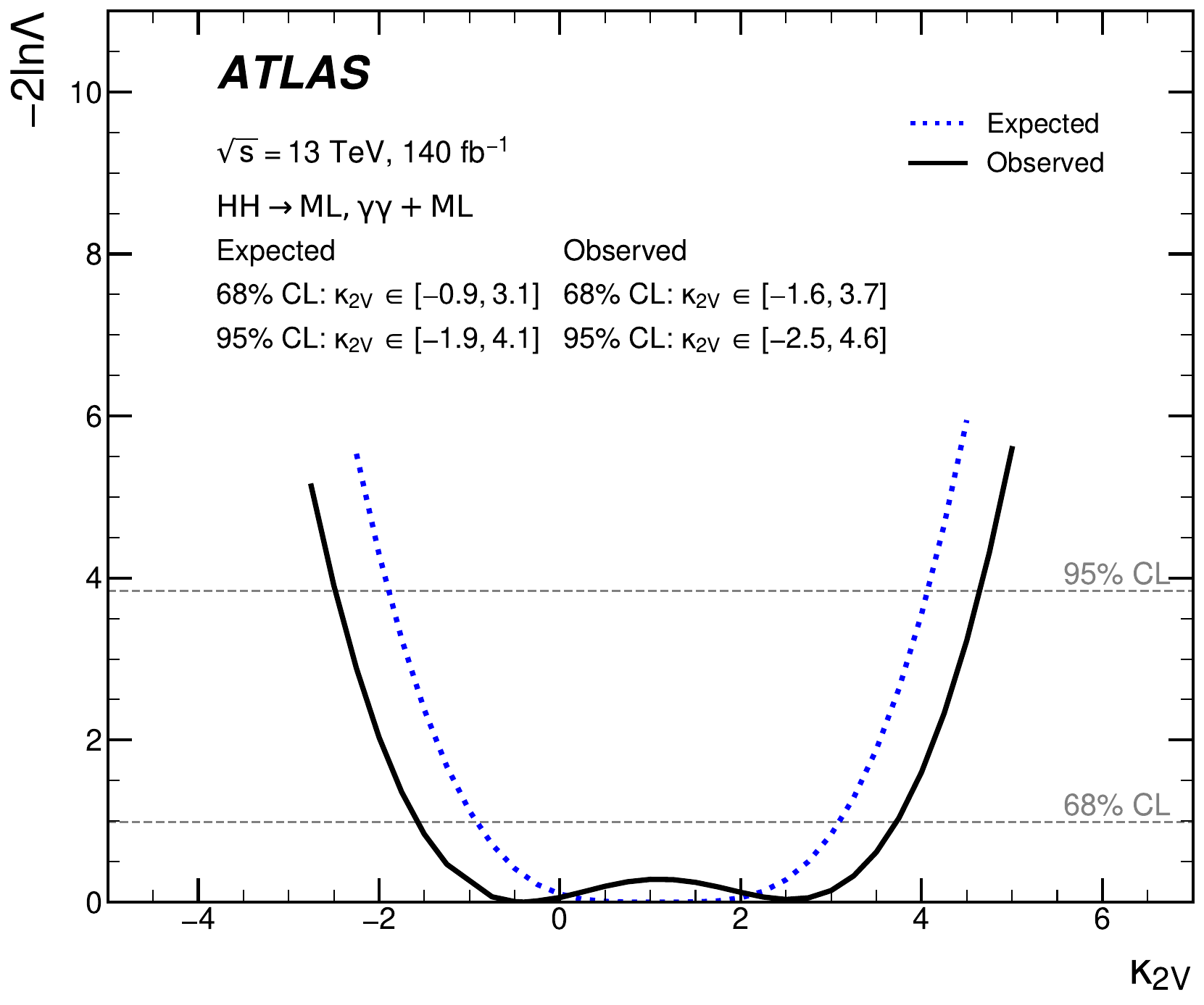}
}
\caption{Observed (solid lines) and expected (dashed line) values of
-2ln$\Lambda$ as a function of (a) \kl, and (b) \kVV.
All other coupling modifiers are fixed to their SM predictions and the
expected limits are computed assuming the SM.
}
\label{fig:likelihood-scans}
\end{figure}

\begin{figure}[!tbhp]
\centering
\subfloat[]{\includegraphics[width=0.45\textwidth]{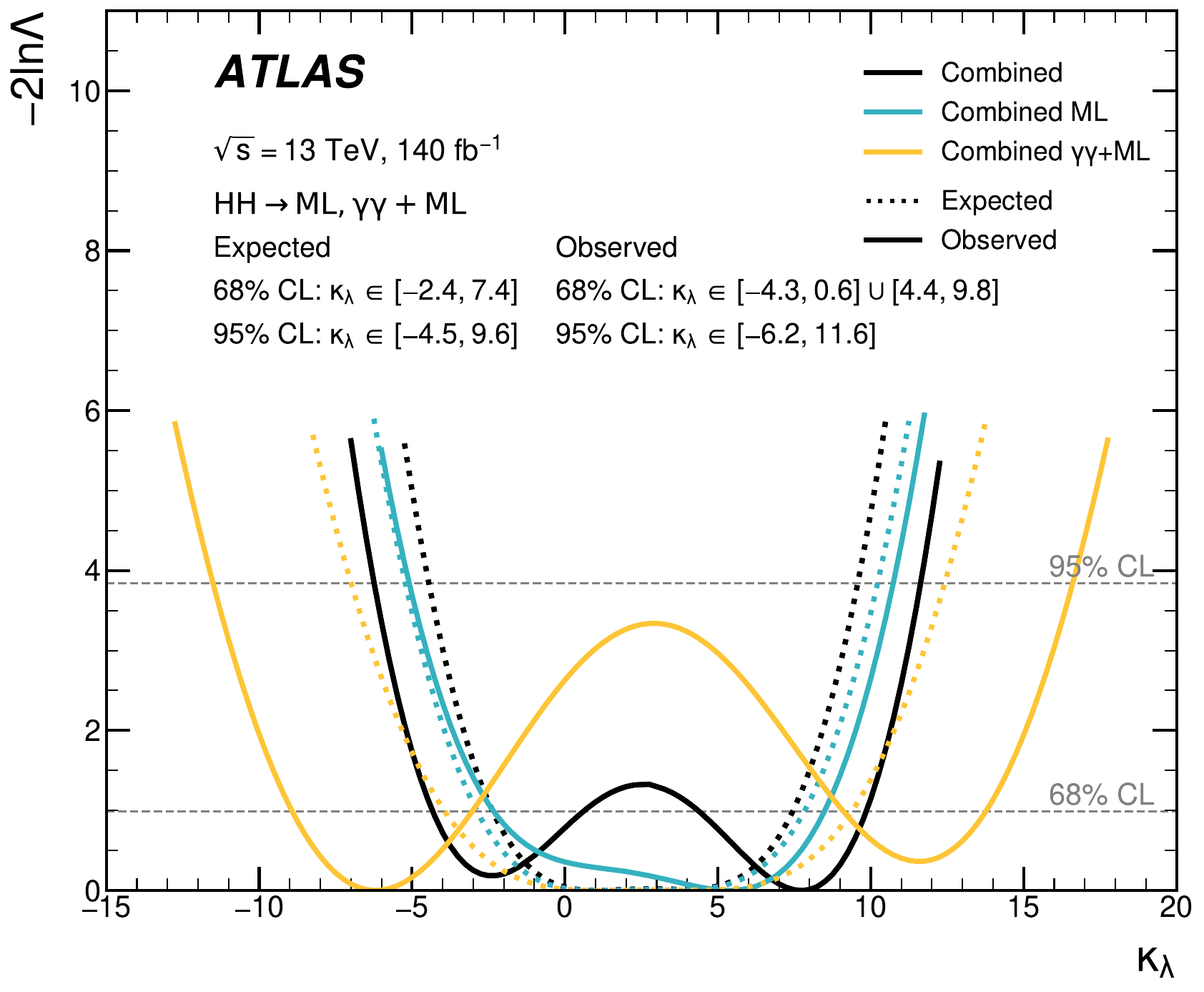}
}
\subfloat[]{\includegraphics[width=0.45\textwidth]{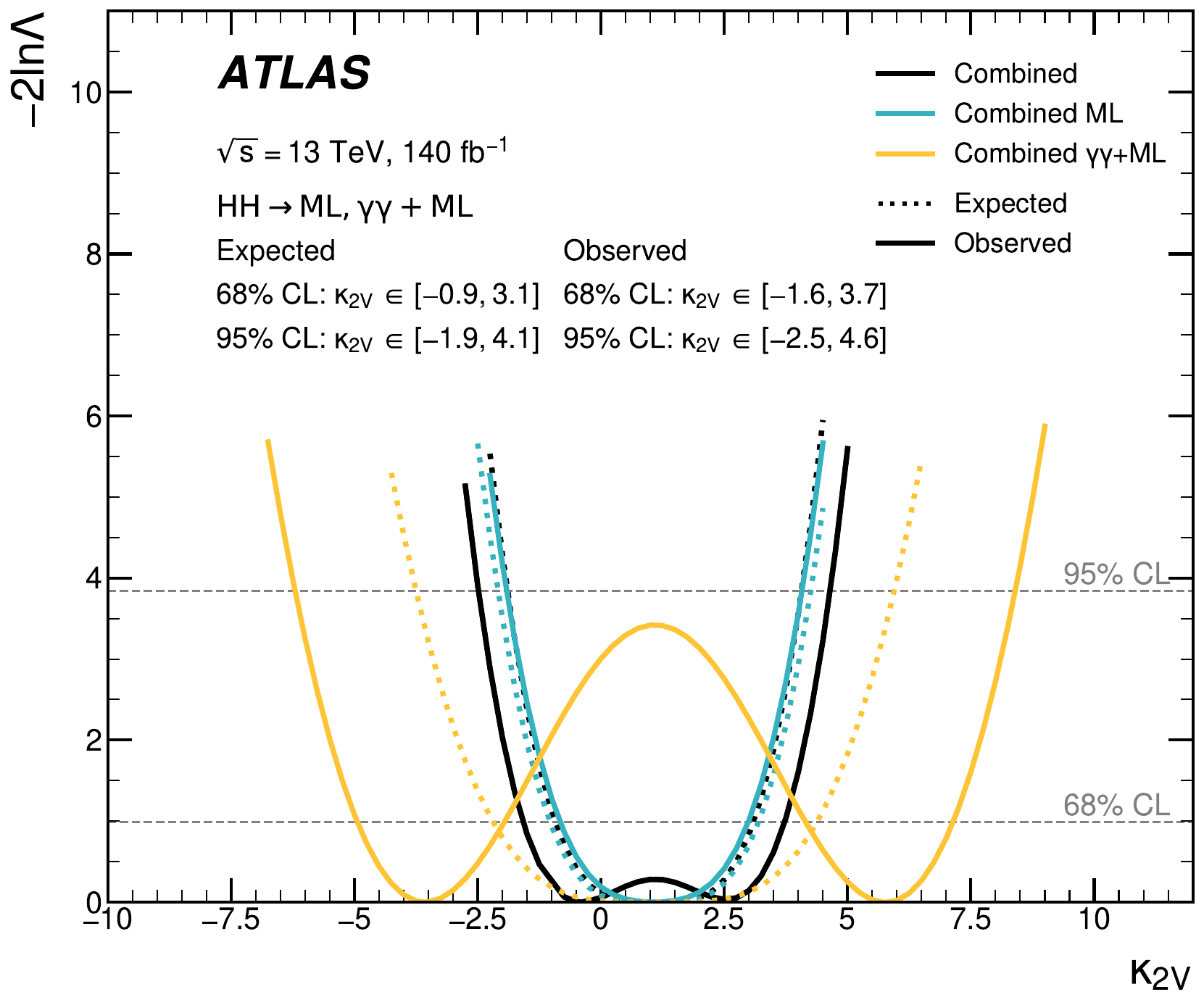}
}
\caption{Observed (solid lines) and expected (dashed line) values of
-2ln$\Lambda$ as a function of (a) \kl, and (b) \kVV\ for the ML
channels, the \yyML\ channels, and their combination.  All other
coupling modifiers are fixed to their SM predictions and the
expected limits are computed assuming the SM.
}
\label{fig:likelihood-scans-split}
\end{figure}


\FloatBarrier
 
\section{Conclusion}

A search for \HH\ production targeting the $bbZZ$, 4$V$,
$VV\tau\tau$, 4$\tau$, $\gamma\gamma VV$\ and $\gamma\gamma\tau\tau$\
decay channels is performed for the first time in ATLAS\@.  Final states
are categorised based on the multiplicity of light charged leptons
(electrons or muons), hadronically decaying tau leptons, and photons.
BDTs are used to separate signal from backgrounds in eight of the nine
explored channels.  The main background processes to the ML channels
involving vector bosons and top-quarks are estimated from MC
simulation and normalised to data.  Background processes involving
charge-misidentification of leptons, non-prompt leptons,
misidentification of hadronic tau leptons, and non-resonant \yy\ production
are estimated by using data-driven methods.  The analysis is performed
with proton--proton collision data at \sqrts~=~13\,\TeV\ collected
from 2015 to 2018 with the ATLAS detector at the LHC, corresponding to
an integrated luminosity of 140\,\ifb.
 
Observed (expected) limits of \muHHobs\ (\muHHexpzero) times the SM
prediction are set on the \HH\ signal strength, under the
background-only hypothesis.  The self-coupling strength modifier, \kl,
is observed (expected) to be constrained to be
$\klobsdown \leq \kl \leq \klobsup$\
($\klexpdown \leq \kl \leq \klexpup$) and \kVV\ is observed (expected)
to be constrained to be $\kvvobsdown \leq \kVV \leq \kvvobsup$\
($\kvvexpdown \leq \kVV \leq \kvvexpup$), all at 95\% CL with other
couplings except for the one being probed fixed to their SM values,
and assuming the SM for the expected limits.  The sensitivity of the
results in all channels is limited by the statistical precision on the
available data.  The results presented in this study have comparable
sensitivity to the other channels already investigated by ATLAS and
CMS, and will contribute to improve the global sensitivity to \HH\
production.


\section*{Acknowledgements}


We thank CERN for the very successful operation of the LHC and its injectors, as well as the support staff at
CERN and at our institutions worldwide without whom ATLAS could not be operated efficiently.
 
The crucial computing support from all WLCG partners is acknowledged gratefully, in particular from CERN, the ATLAS Tier-1 facilities at TRIUMF/SFU (Canada), NDGF (Denmark, Norway, Sweden), CC-IN2P3 (France), KIT/GridKA (Germany), INFN-CNAF (Italy), NL-T1 (Netherlands), PIC (Spain), RAL (UK) and BNL (USA), the Tier-2 facilities worldwide and large non-WLCG resource providers. Major contributors of computing resources are listed in Ref.~\cite{ATL-SOFT-PUB-2023-001}.
 
We gratefully acknowledge the support of ANPCyT, Argentina; YerPhI, Armenia; ARC, Australia; BMWFW and FWF, Austria; ANAS, Azerbaijan; CNPq and FAPESP, Brazil; NSERC, NRC and CFI, Canada; CERN; ANID, Chile; CAS, MOST and NSFC, China; Minciencias, Colombia; MEYS CR, Czech Republic; DNRF and DNSRC, Denmark; IN2P3-CNRS and CEA-DRF/IRFU, France; SRNSFG, Georgia; BMBF, HGF and MPG, Germany; GSRI, Greece; RGC and Hong Kong SAR, China; ISF and Benoziyo Center, Israel; INFN, Italy; MEXT and JSPS, Japan; CNRST, Morocco; NWO, Netherlands; RCN, Norway; MNiSW, Poland; FCT, Portugal; MNE/IFA, Romania; MESTD, Serbia; MSSR, Slovakia; ARIS and MVZI, Slovenia; DSI/NRF, South Africa; MICIU/AEI, Spain; SRC and Wallenberg Foundation, Sweden; SERI, SNSF and Cantons of Bern and Geneva, Switzerland; NSTC, Taipei; TENMAK, T\"urkiye; STFC/UKRI, United Kingdom; DOE and NSF, United States of America.
 
Individual groups and members have received support from BCKDF, CANARIE, CRC and DRAC, Canada; CERN-CZ, FORTE and PRIMUS, Czech Republic; COST, ERC, ERDF, Horizon 2020, ICSC-NextGenerationEU and Marie Sk{\l}odowska-Curie Actions, European Union; Investissements d'Avenir Labex, Investissements d'Avenir Idex and ANR, France; DFG and AvH Foundation, Germany; Herakleitos, Thales and Aristeia programmes co-financed by EU-ESF and the Greek NSRF, Greece; BSF-NSF and MINERVA, Israel; Norwegian Financial Mechanism 2014-2021, Norway; NCN and NAWA, Poland; La Caixa Banking Foundation, CERCA Programme Generalitat de Catalunya and PROMETEO and GenT Programmes Generalitat Valenciana, Spain; G\"{o}ran Gustafssons Stiftelse, Sweden; The Royal Society and Leverhulme Trust, United Kingdom.
 
In addition, individual members wish to acknowledge support from CERN: European Organization for Nuclear Research (CERN PJAS); Chile: Agencia Nacional de Investigaci\'on y Desarrollo (FONDECYT 1190886, FONDECYT 1230812, FONDECYT 1230987); China: Chinese Ministry of Science and Technology (MOST-2023YFA1605700), National Natural Science Foundation of China (NSFC - 12175119, NSFC 12275265, NSFC-12075060); Czech Republic: Czech Science Foundation (GACR - 24-11373S), Ministry of Education Youth and Sports (FORTE CZ.02.01.01/00/22\_008/0004632), PRIMUS Research Programme (PRIMUS/21/SCI/017); EU: H2020 European Research Council (ERC - 101002463); European Union: European Research Council (ERC - 948254, ERC 101089007), Horizon 2020 Framework Programme (MUCCA - CHIST-ERA-19-XAI-00), European Union, Future Artificial Intelligence Research (FAIR-NextGenerationEU PE00000013), Italian Center for High Performance Computing, Big Data and Quantum Computing (ICSC, NextGenerationEU); France: Agence Nationale de la Recherche (ANR-20-CE31-0013, ANR-21-CE31-0013, ANR-21-CE31-0022, ANR-22-EDIR-0002), Investissements d'Avenir Labex (ANR-11-LABX-0012); Germany: Baden-Württemberg Stiftung (BW Stiftung-Postdoc Eliteprogramme), Deutsche Forschungsgemeinschaft (DFG - 469666862, DFG - CR 312/5-2); Italy: Istituto Nazionale di Fisica Nucleare (ICSC, NextGenerationEU), Ministero dell'Università e della Ricerca (PRIN - 20223N7F8K - PNRR M4.C2.1.1); Japan: Japan Society for the Promotion of Science (JSPS KAKENHI JP21H05085, JSPS KAKENHI JP22H01227, JSPS KAKENHI JP22H04944, JSPS KAKENHI JP22KK0227); Netherlands: Netherlands Organisation for Scientific Research (NWO Veni 2020 - VI.Veni.202.179); Norway: Research Council of Norway (RCN-314472); Poland: Ministry of Science and Higher Education (IDUB AGH, POB8, D4 no 9722), Polish National Agency for Academic Exchange (PPN/PPO/2020/1/00002/U/00001), Polish National Science Centre (NCN 2021/42/E/ST2/00350, NCN OPUS nr 2022/47/B/ST2/03059, NCN UMO-2019/34/E/ST2/00393, UMO-2020/37/B/ST2/01043, UMO-2021/40/C/ST2/00187, UMO-2022/47/O/ST2/00148, UMO-2023/49/B/ST2/04085); Slovenia: Slovenian Research Agency (ARIS grant J1-3010); Spain: Generalitat Valenciana (Artemisa, FEDER, IDIFEDER/2018/048), Ministry of Science and Innovation (MCIN \& NextGenEU PCI2022-135018-2, MICIN \& FEDER PID2021-125273NB, RYC2019-028510-I, RYC2020-030254-I, RYC2021-031273-I, RYC2022-038164-I), PROMETEO and GenT Programmes Generalitat Valenciana (CIDEGENT/2019/023, CIDEGENT/2019/027); Sweden: Swedish Research Council (Swedish Research Council 2023-04654, VR 2018-00482, VR 2022-03845, VR 2022-04683, VR 2023-03403, VR grant 2021-03651), Knut and Alice Wallenberg Foundation (KAW 2018.0157, KAW 2018.0458, KAW 2019.0447, KAW 2022.0358); Switzerland: Swiss National Science Foundation (SNSF - PCEFP2\_194658); United Kingdom: Leverhulme Trust (Leverhulme Trust RPG-2020-004), Royal Society (NIF-R1-231091); United States of America: U.S. Department of Energy (ECA DE-AC02-76SF00515), Neubauer Family Foundation.


\clearpage
 
\begin{DIFnomarkup}
\appendix
\part*{Appendix}
\addcontentsline{toc}{part}{Appendix}

The BDT input variables used in the different ML
channels are summarised in Tables~\ref{tab:ml_bdtvars_4l_3l_ss2l} and
\ref{tab:ml_bdtvars_sslltau_lltautau_ltautau} and in the \yyl\ and \yytau\
channels in Table~\ref{tab:ml_bdtvars_yyml}.
 
\begin{DIFnomarkup}
\begin{table}[!tbhp]
\begin{center}
\caption{Variables used as inputs to the \bbllll, \lll, and \SSll\ channel BDTs.  The indices $i$\ and $j$\ refer to the indices of the \pt-ordered objects.}
{\small
\begin{tabular}{ l l c c c} \toprule
Variable                     & Description                                                     & \bbllll     & \lll             & \SSll     \\ \toprule
$\pt(\ell_i)$                & \pt\ of the $i$th lepton                                        & $i=1,2,3,4$ & --               & --         \\
$|\eta(\ell_i)|$             & Absolute $\eta$\ of the $i$th lepton                            & $i=1,2,3,4$ & --               & $i=1,2$    \\
$\etcone(\ell_i)$            & Isolation metric (where $E_{\text{T}}^{\Delta R < 0.3}$\ = total transverse  & $i=1,2,3,4$ & --      & --         \\
& energy deposited in a cone of radius $R=0.3$\ around            &             &                  &            \\
& the lepton, and $E_{\text{T}} =$\ lepton transverse energy)      &             &                  &            \\
Dilepton type                & $\mu\mu = 1$, $e\mu$/$\mu e = 2$, $ee = 3$                      & --          & --               & \cmark     \\
\mlilj                       & Invariant mass of the $i$th and $j$th leptons                   & --          & $i,j=1,2$        & $i,j=1,2$  \\
&                                                                 &             & $i,j=1,3$        &            \\
&                                                                 &             & $i,j=2,3$        &            \\ \\[-1em]
\mllonshell                  & Invariant mass of pair of SFOC leptons that                     & \cmark      & \cmark           & --         \\
& minimises the difference with the $Z$ boson mass                &             &                  &            \\
\mlloffshell                 & Invariant mass of the other SFOC lepton pair                    & \cmark      & --               & --         \\ \\[-1em]
min. \mllSFOS                & Minimum invariant mass out of all SFOC pairs                    & --          & \cmark           & --         \\
\mllll                       & Invariant mass of four leptons                                  & \cmark      & --               & --         \\
\mlll                        & Invariant mass of three leptons                                 & --          & \cmark           & --         \\
\mlclosej                    & Invariant mass of the $i$th lepton and its closest jet          & --          & $i=1,2,3$        & $i=1,2$    \\
\mllljj                      & Invariant mass of the three leptons and the leading             & --          & \cmark           & --         \\
& (or two leading, for events with $\Njet \ge 2$) jets            &             &                  &            \\
\mjj                         & Invariant mass of the two leading jets                          & \cmark      & --               & --         \\
\mall                        & Invariant mass of all selected objects in the event             & --          & --               & \cmark     \\
\mtw                         & Transverse mass of the $i$th lepton and the \met                & --          & --               & $i=1,2$    \\
\detaloneltwo                & Separation in $\eta$\ between the first and second leptons      & --          & --               & \cmark     \\
\dRlilj                      & Separation in $R$\ between the $i$th and $j$th leptons          & --          & $i,j=1,2$        & $i,j=1,2$  \\
&                                                                 &             & $i,j=1,3$        &           \\
&                                                                 &             & $i,j=2,3$        &           \\
\dRlclosej                   & Separation in $R$\ between the $i$th lepton and its             & --          & $i=1,2,3$        & $i=1,2$    \\
& closest jet                                                     &             &                  &            \\
min. \dRljet                 & Minimum separation in $R$\ between any lepton and any jet       & --          & --               & \cmark     \\
\mylt                        & Scalar sum of the \pt\ of all leptons and the \met              & --          & \cmark           & \cmark     \\
\myht                        & Scalar sum of the \pt of all jets                               & --          & \cmark           & \cmark     \\
\myst                        & Scalar sum of the \pt\ of all objects in the event              & \cmark      & \cmark           & --         \\
\sumQ                        & Sum of all lepton charges                                       & --          & --               & \cmark     \\
\Njet                        & Number of jets in the event                                     & --          & --               & \cmark     \\
\Nbjet                       & Number of $b$-jets in the event                                 & \cmark      & --               & --         \\
\ptjone                      & \pt\ of the leading jet                                         & \cmark      & --               & --         \\
\ptjj                        & \pt\ of the leading dijet system                                & \cmark      & --               & --         \\
\met                         & Magnitude of the missing transverse momentum                    & \cmark      & \cmark           & \cmark     \\
\dphimetj                    & $\phi$ angle between the \met\ and the leading jet              & \cmark      & --               & --         \\ \bottomrule
\end{tabular}
}
\label{tab:ml_bdtvars_4l_3l_ss2l}
\end{center}
\end{table}
\end{DIFnomarkup}
 
\begin{DIFnomarkup}
\begin{table}[!tbhp]
\begin{center}
\caption{Variables used as inputs to the \SSlltau, \lltautau, and \ltautau\ channel BDTs.  The indices $i$\ and $j$\ refer to the indices of the \pt-ordered objects.}
{\small
\begin{tabular}{ l l c c c} \toprule
Variable                      & Description                                                      & \SSlltau    & \lltautau     & \ltautau    \\ \toprule
Dilepton type                & $\mu\mu = 1$, $e\mu$/$\mu e = 2$, $ee = 3$                       & --          & \cmark        & --          \\
\mloneltwo                    & Invariant mass of the two leptons                                & --          & \cmark        & --          \\
\mloneclosej                  & Invariant mass of the leading lepton and its closest jet         & \cmark      & --            & \cmark      \\
\mlj                          & Invariant mass of the $i$th lepton and $j$th jet                 & $i,j=1,1$   & --            & --          \\
&                                                                  & $i,j=1,2$   &               &             \\
&                                                                  & $i,j=2,1$   &               &             \\
\detall                       & Separation in $\eta$\ between the two leptons                    & \cmark      & --            & --          \\
\dRll                         & Separation in $R$\ between the two leptons                       & \cmark      & \cmark        & --          \\
\dRlijj                       & Separation in $R$\ between the $i$th lepton and $j$th jet        & $i,j=1,1$   & --            & $i,j=1,1$   \\
&                                                                  &             &               & $i,j=1,2$   \\
\dRlclosej                    & Separation in $R$\ between the $i$th lepton and its              & $i=1,2$     & --            & --          \\
& closest jet                                                      &             &               &             \\
\ptjone                       & \pt\ of the leading jet                                          & --          & --            & \cmark      \\
\met                          & Magnitude of the missing transverse momentum                     & --          & --            & \cmark      \\ \\[-1em]
\boosttauji                   & Polar angle between the \tauhadvis\ and the $i$th jet after a    & $i=1,2$     & --            & --          \\
& Lorentz boost to the dilepton system                             &             &               &             \\ \\[-1em]
\boostlijj                    & Separation in $R$ between the $i$th lepton and $j$th jet         & $i,j=1,2$   & --            & --          \\
& after a Lorentz boost to the \tauhadvis\ and $i$th lepton system & $i,j=2,1$   &               &             \\ \\[-1em]
\mtautau                      & Invariant mass of the two \tauhadvis\                            & --          & \cmark        & \cmark      \\
\dRltwotauone                 & Separation in $R$ between the second lepton and first \tauhadvis & --          & \cmark        & --          \\
\dRlonetautau                 & Separation in $R$ between the first lepton and the               & --          & \cmark        & \cmark      \\
& di-\tauhadvis\ system                                            &             &               &             \\
\mltwotauone                  & Invariant mass of the second lepton and first \tauhadvis         & --          & \cmark        & --          \\
\mltautau                     & Invariant mass of the lepton and two \tauhadvis\                 & --          & --            & \cmark      \\
\vecsumptlj                   & Vector sum of the $\pt$ of the lepton and its closest jet        & --          & --            & \cmark      \\
\vecsumpttautau               & Vector sum of the $\pt$ of the two \tauhadvis\                   & --          & \cmark        & \cmark      \\  \bottomrule
\end{tabular}
}
\label{tab:ml_bdtvars_sslltau_lltautau_ltautau}
\end{center}
\end{table}
\end{DIFnomarkup}
 
\begin{DIFnomarkup}
\begin{table}[!tbhp]
\begin{center}
\caption{Variables used as inputs to the \yyl\ and \yytau\ channel BDTs.  Photons and jets are \pt\ ordered.}
{\small
\begin{tabular}{ l l c c c } \toprule
Variable                                 &  Description                                           & \yyl    & \yytau         \\ \toprule
$\pt(\yy)$                               & \pt\ of the diphoton system                            & \cmark  & \cmark         \\
$\pt(\ell)$                              & \pt\ of the lepton                                     & \cmark  & --             \\
$\pt(\tauhvis)$                          & \pt\ of the \tauhvis                                   & --      & \cmark          \\
\met                                     & Magnitude of the missing transverse momentum           & \cmark  & \cmark         \\
$\phi(\met)$                             & $\phi$\ direction of the \met                          & --      & \cmark         \\
$\eta(\ell\met)$                         & $\eta$ of the lepton-\met\ system                      & \cmark  & --             \\
$\eta(\gamma_1)$                         & $\eta$ of the leading photon                           & --      & \cmark         \\
$N_{\text{central-jets}}$                   & Number of jets with $|\eta| < 2.5$                     & \cmark  & \cmark         \\
$\Delta R(\ell, \met)$                   & \dR\ between the lepton and the \met                   & \cmark  & --             \\
$\Delta R(\yy, \ell\met)$                & \dR\ between the diphoton system and the               & \cmark  & --             \\
& lepton-\met\ system                                    &         &                \\
$\Delta\phi(\ell/\tauhvis,\yy)$          & Separation in $\phi$\ between the lepton               & \cmark  & \cmark         \\
& or \tauhvis\ and the diphoton system                   &         &                \\
$\Delta\phi(\gamma_1, \yy)$              & Separation in $\phi$\ between the leading              & \cmark  & \cmark         \\
& photon and the diphoton system                         &         &                \\
min. $\Delta \phi (\met,\text{j},\ell)$  & Minimum $\phi$\ angle between any pair of              & \cmark  & --             \\
& the \met, the lepton, and any jet                      &         &                \\
$\Delta \phi(\met, \yy)$                 & Separation in $\phi$ between the \met and the          & \cmark  & \cmark         \\
& diphoton system                                        &         &                \\ \bottomrule
\end{tabular}
}
\label{tab:ml_bdtvars_yyml}
\end{center}
\end{table}
\end{DIFnomarkup}


\end{DIFnomarkup}
 
\FloatBarrier

\printbibliography
 
\clearpage
 
\begin{flushleft}
\hypersetup{urlcolor=black}
{\Large The ATLAS Collaboration}

\bigskip

\AtlasOrcid[0000-0002-6665-4934]{G.~Aad}$^\textrm{\scriptsize 104}$,
\AtlasOrcid[0000-0001-7616-1554]{E.~Aakvaag}$^\textrm{\scriptsize 17}$,
\AtlasOrcid[0000-0002-5888-2734]{B.~Abbott}$^\textrm{\scriptsize 123}$,
\AtlasOrcid[0000-0002-0287-5869]{S.~Abdelhameed}$^\textrm{\scriptsize 119a}$,
\AtlasOrcid[0000-0002-1002-1652]{K.~Abeling}$^\textrm{\scriptsize 56}$,
\AtlasOrcid[0000-0001-5763-2760]{N.J.~Abicht}$^\textrm{\scriptsize 50}$,
\AtlasOrcid[0000-0002-8496-9294]{S.H.~Abidi}$^\textrm{\scriptsize 30}$,
\AtlasOrcid[0009-0003-6578-220X]{M.~Aboelela}$^\textrm{\scriptsize 45}$,
\AtlasOrcid[0000-0002-9987-2292]{A.~Aboulhorma}$^\textrm{\scriptsize 36e}$,
\AtlasOrcid[0000-0001-5329-6640]{H.~Abramowicz}$^\textrm{\scriptsize 154}$
% ...
\bigskip
\\

% ...
$^{17}$Department for Physics and Technology, University of Bergen, Bergen; Norway.\\
% ...
$^{30}$Physics Department, Brookhaven National Laboratory, Upton NY; United States of America.\\
% ...
$^{36}$$^{(a)}$Facult\'e des Sciences Ain Chock, Universit\'e Hassan II de Casablanca;$^{(b)}$Facult\'{e} des Sciences, Universit\'{e} Ibn-Tofail, K\'{e}nitra;$^{(c)}$Facult\'e des Sciences Semlalia, Universit\'e Cadi Ayyad, LPHEA-Marrakech;$^{(d)}$LPMR, Facult\'e des Sciences, Universit\'e Mohamed Premier, Oujda;$^{(e)}$Facult\'e des sciences, Universit\'e Mohammed V, Rabat;$^{(f)}$Institute of Applied Physics, Mohammed VI Polytechnic University, Ben Guerir; Morocco.\\
% ...
$^{45}$Physics Department, Southern Methodist University, Dallas TX; United States of America.\\
% ...
$^{50}$Fakult\"{a}t Physik , Technische Universit{\"a}t Dortmund, Dortmund; Germany.\\
% ...
$^{56}$II. Physikalisches Institut, Georg-August-Universit\"{a}t G\"ottingen, G\"ottingen; Germany.\\
% ...
$^{104}$CPPM, Aix-Marseille Universit\'e, CNRS/IN2P3, Marseille; France.\\
% ...
$^{119}$$^{(a)}$New York University Abu Dhabi, Abu Dhabi;$^{(b)}$United Arab Emirates University, Al Ain; United Arab Emirates.\\
% ...
$^{123}$Homer L. Dodge Department of Physics and Astronomy, University of Oklahoma, Norman OK; United States of America.\\
% ...
$^{154}$Raymond and Beverly Sackler School of Physics and Astronomy, Tel Aviv University, Tel Aviv; Israel.\\
% ...
$^{a}$ Also Affiliated with an institute covered by a cooperation agreement with CERN.\\
$^{b}$ Also at An-Najah National University, Nablus; Palestine.\\
$^{c}$ Also at Borough of Manhattan Community College, City University of New York, New York NY; United States of America.\\
$^{d}$ Also at Center for Interdisciplinary Research and Innovation (CIRI-AUTH), Thessaloniki; Greece.\\
$^{e}$ Also at Centro Studi e Ricerche Enrico Fermi; Italy.\\
$^{f}$ Also at CERN, Geneva; Switzerland.\\
% ...

\end{flushleft}


\end{document}